\documentclass[12pt]{iopart}

\usepackage{graphicx}
\usepackage{amssymb}
\usepackage{iopams}

\newcommand{\dfrac}[2]{\frac{\displaystyle #1}{\displaystyle #2}}
\newcommand{\email}[1]{\ead{#1}}
\newcommand{\affiliation}[1]{\address{#1}}
\newcommand{\sss}[1]{{\scriptscriptstyle{#1}}}

\newcommand{\ddprime}[1]{\mathaccent"707D{#1}}

\newcommand{\heaviside}[1]{\mathrm{H}\!\left( #1 \right)}
\newcommand{\vect}[1]{\boldsymbol{#1}}
\newcommand{\lvev}[1]{\left\langle#1 \right\rangle}
\newcommand{\vev}[1]{\langle#1 \rangle}
\newcommand{\window}[2]{W_{\negthinspace #1}\negthinspace\left(#2\right)}
\newcommand{\FFTheta}[1]{\hat{\Theta}_{{#1}}}
\newcommand{\gammain}[2]{\gamma_\un\negthinspace\left(#1,#2\right)}

\newcommand{\MeV}{\mathrm{MeV}}

\newcommand{\Area}{\mathcal{A}}
\newcommand{\calM}{\mathcal{M}}
\newcommand{\calL}{\mathcal{L}}

\newcommand{\uh}{\mathrm{h}}
\newcommand{\ub}{\mathrm{b}}
\newcommand{\ue}{\mathrm{e}}
\newcommand{\un}{\mathrm{n}}
\newcommand{\ud}{\mathrm{d}}

\newcommand{\up}{\mathrm{p}}
\newcommand{\uv}{\mathrm{v}}
\newcommand{\uc}{\mathrm{c}}
\newcommand{\ur}{\mathrm{r}}
\newcommand{\ui}{\mathrm{i}}
\newcommand{\uuc}{\mathrm{uc}}

\newcommand{\uiso}{\ell\ell\theta}
\newcommand{\ueq}{\mathrm{eq}}
\newcommand{\umat}{\mathrm{mat}}
\newcommand{\urad}{\mathrm{rad}}
\newcommand{\uCMB}{\mathrm{\sss{CMB}}}

\newcommand{\ufov}{\mathrm{fov}}

\newcommand{\etav}{\eta_\uv}

\newcommand{\calH}{\mathcal{H}}
\newcommand{\calS}{\mathcal{S}}
\newcommand{\calD}{\mathcal{D}}

\newcommand{\calG}{\mathcal{G}}
\newcommand{\calP}{\mathcal{P}}
\newcommand{\scaling}{\calS}

\newcommand{\Tb}{\bar{T}}
\newcommand{\nablab}{\bar{\nabla}}
\newcommand{\mub}{\bar{\mu}}
\newcommand{\nub}{\bar{\nu}}

\newcommand{\Xdd}{\ddot{X}}
\newcommand{\Xd}{\dot{X}}

\newcommand{\Xp}{\acute{X}}
\newcommand{\bk}{\vect{k}}
\newcommand{\bu}{\vect{u}}
\newcommand{\vecu}{\vec{u}}

\newcommand{\bX}{\vect{X}}
\newcommand{\bx}{\vect{x}}

\newcommand{\bXd}{\dot{\bX}}
\newcommand{\bXp}{\acute{\bX}}
\newcommand{\bXdd}{\ddot{\bX}}
\newcommand{\bXpp}{\ddprime{\bX}}

\newcommand{\vecX}{\vec{X}}
\newcommand{\vecXd}{\dot{\vecX}}
\newcommand{\vecXp}{\acute{\vecX}}
\newcommand{\vecXdd}{\ddot{\vecX}}

\newcommand{\bp}{\vec{p}}

\newcommand{\bq}{\vec{q}}

\newcommand{\Pe}{P_\ue}
\newcommand{\corrini}{\ell_\uc}
\newcommand{\resini}{\ell_\ur}
\newcommand{\ellinf}{\ell_\infty}
\newcommand{\rhoinf}{\rho_\infty}
\newcommand{\rholoop}{\rho_\circ}

\newcommand{\horizon}{d_\uh}
\newcommand{\Nppcl}{N_{\mathrm{ppcl}}}
\newcommand{\frachor}{\alpha}
\newcommand{\frachorscal}{\frachor_{\min}}
\newcommand{\frachorlong}{\frachor_\infty}
\newcommand{\frachorini}{\frachor_\uc}
\newcommand{\frachorres}{\frachor_\ur}

\newcommand{\numloop}{n}
\newcommand{\noconst}{C}
\newcommand{\const}{\noconst_\circ}
\newcommand{\power}{p}
\newcommand{\kperp}{l}
\newcommand{\unitn}{\hat{n}}

\newcommand{\Tcmb}{T_\uCMB}
\newcommand{\GHz}{\textrm{GHz}}
\newcommand{\Gpc}{\textrm{Gpc}}
\newcommand{\ang}{\theta}

\newcommand{\angfov}{\ang_\ufov}
\newcommand{\angsqz}{\ang}

\newcommand{\angx}{\alpha}
\newcommand{\angy}{\beta}

\newcommand{\ep}{\epsilon}
\newcommand{\half}{\frac{1}{2}}
\newcommand{\Ga}{\Gamma}
\newcommand{\vb}{\bar{v}}
\newcommand{\tb}{\bar{t}}
\newcommand{\BT}{B}
\newcommand{\ka}{\kappa}
\newcommand{\alphac}{c_1}
\newcommand{\betac}{c_2}
\newcommand{\corr}{\hat{\xi}}
\newcommand{\chimax}{\Lambda}
\newcommand{\why}{Y}
\newcommand{\fNL}{f_{\mathrm{NL}}}
\newcommand{\tauNL}{\tau_{\mathrm{NL}}}

\begin{document}

\title[Cosmic strings and their induced non-Gaussianities]{Cosmic
  strings and their induced non-Gaussianities in the cosmic microwave
  background}

\author{Christophe Ringeval}
\email{christophe.ringeval@uclouvain.be}
\affiliation{Institute of Mathematics and Physics, Centre for
  Cosmology, Particle Physics and Phenomenology, Louvain University, 2
  Chemin du Cyclotron, 1348 Louvain-la-Neuve, Belgium} \date{today}

\begin{abstract}
  Motivated by the fact that cosmological perturbations of
  inflationary quantum origin were born Gaussian, the search for
  non-Gaussianities in the cosmic microwave background (CMB)
  anisotropies is considered as the privileged probe of non-linear
  physics in the early universe. Cosmic strings are active sources of
  gravitational perturbations and incessantly produce non-Gaussian
  distortions in the CMB. Even if, on the currently observed angular
  scales, they can only contribute a small fraction of the CMB angular
  power spectrum, cosmic strings could actually be the main source of
  its non-Gaussianities. In this article, after having reviewed the
  basic cosmological properties of a string network, we present the
  signatures Nambu--Goto cosmic strings would induce in various
  observables ranging from the one-point function of the temperature
  anisotropies to the bispectrum and trispectrum. It is shown that
  string imprints are significantly different than those expected from
  the primordial type of non-Gaussianity and could therefore be easily
  distinguished.
\end{abstract}


\section{Motivations}
\label{sec:introduction}

The origin of cosmic strings dates back to the discovery that
cosmological phase transitions triggered by the spontaneous breakdown
of the fundamental interaction symmetries may form topological
defects~\cite{Kirzhnits:1972,Kobsarev:1974, Kibble:1976}. Cosmic
strings belong to the class of line-like topological defects, as
opposed to point-like monopoles and the membrane shaped domain
walls. As shown by Kibble, the appearance of defects in any field
theory is related to the topology of the vacuum
manifold~\cite{Kibble:1976}. If the ground state of a field theory
experiences a spontaneous breakdown from a symmetry group $G$ to a
subgroup $H$, Kibble showed that cosmic strings will be formed if the
first homotopy group $\pi_1(G/H) \neq I$ is non-trivial. In other
words, if non-contractile loops can be found in the manifold
$\calM=G/H$ of equivalent vacua. Similarly, the other homotopy groups
$\pi_0$ and $\pi_2$ determine the formation of domain walls and
monopoles respectively. Once formed and cooled, these defects cannot
be unfolded, precisely due to their non-trivial topological
configuration over the vacuum manifold of the theory. This simple
statement suggests that cosmic strings, and topological defects in
general, are a natural outcome of the unification of the fundamental
interactions in the context of Cosmology. As remnants of unified
forces, their discovery would be an incredible opportunity to probe
extremely high energy physics with ``a telescope''.

In the last thirty years, many works have been devoted to the
cosmological consequences, signatures and searches for topological
defects~\cite{Hindmarsh:1994re, Vilenkin:2000, Sakellariadou:2006qs,
  Peter:1208401}. They have pushed cosmic strings to the privileged
place to be generically compatible with observations. Indeed, domain
walls and monopoles are prone to suffer from the cosmological
catastrophe problem: their formation is sufficiently efficient (or
their annihilation sufficiently inefficient) to either overclose the
universe or spoil the Big-Bang Nucleosynthesis (BBN)
predictions~\cite{1996RPPh.59.1493S, 2005APh.23.313C}. For domain
walls, this implies that either they should be extremely light,
i.e. formed at an energy scale less than a few $\MeV$, or no discrete
symmetry should have been broken during the cooling of the
universe. There is not so much choice for the monopoles: if
interactions were unified, monopoles would have been formed. The
homotopy group of $\pi_2(G/H)$ with $H$ containing the $U(1)$ of
electroweak interactions is indeed non-trivial\footnote{As often with
  topological defects, sensitivity to the underlying model is such
  that one can often find a counter-example of any result. Both of
  these statements, on walls and monopoles, can be evaded in some
  particular models or with some amount of fine-tuning, as for
  instance if cosmic strings can be attached to them and catalyse
  annihilations~\cite{PhysRevD.26.435, PhysRevLett.45.1}.}. Cosmic
inflation was originally designed to solve the monopole problem. If a
phase of accelerated expansion of the universe occurs, then any
defects will be diluted enough to no longer have any (dramatic)
consequences on cosmology~\cite{Guth:1980zm, Starobinsky:1980te,
  Linde:1981mu, Starobinsky:1982ee}. Meanwhile, Inflationary Cosmology
solves the flatness and homogeneity problem of the standard Big-Bang
model, explains the origin and spectrum of the Cosmic Microwave
Background (CMB) anisotropies, as the formation of the large scale
structures~\cite{Mukhanov:1990me, Martin:2006rs}. Inflation provides a
priori an easy solution to the topological defects problem by diluting
them to at most one per Hubble radius. However, one has to keep in
mind that this mechanism works only if the defects were formed before
inflation, and even in that case some may
survive~\cite{2007PhRvD.76h3510A}. This has to be the case for
monopoles and heavy walls, but not for local strings. On the contrary,
exhaustive analysis of particle physics motivated inflationary models,
embedding the Standard Model $SU(3)\times SU(2) \times U(1)$, has
shown that strings are generically produced at the end of
inflation~\cite{Jeannerot:2003qv}. In this picture, our universe
should contain cosmic strings whose properties are closely related to
those of the inflaton~\cite{Rocher:2004my, Rocher:2006nh,
  Battye:2006pk}. String Theory provides an alternative framework to
Field Theories: brane inflationary models propose that the accelerated
expansion of the universe is induced by the motion of branes in warped
and compact extra-dimensions~\cite{Becker:2007, Dvali:1998pa,
  Alexander:2001ks, Kachru:2003sx}. Inflation ends when two branes
collide and such a mechanism again triggers the formation of
one-dimensional cosmological extended objects, dubbed cosmic
superstrings~\cite{Burgess:2001fx, Sarangi:2002yt, Dvali:2003zj,
  Jones:2003da}. These objects may be cosmologically stretched
fundamental strings or one-dimensional D-brane~\cite{Davis:2005,
  Copeland:2009ga}. Although cosmic superstrings are of a different
nature than their topological analogue, they produce the same
gravitational effects and share similar cosmological
signatures~\cite{Sakellariadou:2008ie, Sakellariadou:2009ev}.

Among the expected signatures, cosmic strings induce temperature
anisotropies in the CMB with an amplitude typically given by $GU$,
where $U$ is the string energy per unit length\footnote{To avoid any
  confusion with Greek tensor indices, we will use the Carter's
  notations $U$ and $T$ for the string energy density and
  tension~\cite{Carter:1989dp}.} and $G$ the Newton
constant~\cite{Bouchet:1988}. For the Grand Unified Theory (GUT)
energy scale, one has $GU \simeq 10^{-5}$, which precisely corresponds
to the observed amplitude of the CMB temperature
fluctuations~\cite{Durrer:2001cg}. However, the power spectra do not
match: topological defects are active sources of gravitational
perturbations, i.e. they produce perturbations all along the universe
history, and cannot produce the characteristic coherent patterns of
the acoustic peaks~\cite{Durrer:1996, Magueijo:1996px, Albrecht:1997,
  Spergel:2006hy, Komatsu:2008hk}. Current CMB data analyses including
a string contribution suggest that they can only contribute to at most
$10\%$ of the overall anisotropies on the observed angular
scales~\cite{Bouchet:2000hd, Fraisse:2007}. For Abelian cosmic strings
(see Sec.~\ref{sec:kinds}), numerical simulations in
Friedmann--Lema\^{\i}tre--Robertson--Walker (FLRW) spacetimes show
that this corresponds to an upper two-sigma bound $GU < 7\times
10^{-7}$~\cite{Hindmarsh:2007}. Direct detection searches provide less
stringent limits but are applicable to all cosmic string models: $GU<
4\times 10^{-6}$~\cite{Gilbert:1995de, Jeong:2007,
  Amsel:2007ki}. Detecting cosmic strings in the CMB certainly
requires one to go further than the power
spectrum~\cite{Gangui:2001fr, Pogosian:2007gi} (see however
Sec.~\ref{sec:cls}). In fact, strings induce line-like discontinuities
in the CMB temperature through the so-called Gott--Kaiser--Stebbins
effect, which are intrinsically of non-Gaussian
nature~\cite{Gott:1984ef,Kaiser:1984iv}. In the inflationary picture,
cosmological perturbations find their origin in the quantum
fluctuations of the field--metric system, and therefore were born
generically Gaussian. Non-Gaussianities can nevertheless appear from
non-linear effects during inflation or from couplings to other fields
(see the other articles in this issue). These non-Gaussianities are of
the primordial type, i.e. they exist before the cosmological
perturbations reenter the Hubble radius. On the other hand, cosmic
strings are a source of non-Gaussianity at all times and, as we will
see, produce different signals from the CMB point of
view\footnote{Notice that second order perturbations, being
  non-linear, actively generate non-Gaussianities but at a relatively
  small amplitude~\cite{Sefusatti:2007ih, Pitrou:2008ak, Nitta:2009jp,
    Boubekeur:2009uk}.}.

In this article, we review the non-Gaussian features a cosmological
network of cosmic strings produce in the CMB anisotropies. In a first
section, we briefly scan various cosmic string models and emphasize
their similarities and differences for cosmology. Making observable
predictions for cosmic strings faces the problem of understanding
their cosmological evolution. Not only one has to solve the local
dynamics in curved space, but as extended objects, cosmic strings
follow a globally non-local evolution: the fate of one string depends
on its interactions with the others. The cosmological evolution of a
network of cosmic strings is a non-trivial problem which can be
overcome by means of numerical simulations. These simulations permit
an estimation of the various statistical properties affecting the
observational signatures, such as the number of strings per Hubble
radius, their shapes, velocities or the loop density
distribution. Latest results in this area, for the Nambu--Goto (NG)
type of cosmic strings, are presented in Sec.~\ref{sec:evol}. Once the
statistical properties of a cosmological cosmic strings network are
known, it is possible to extract meaningful observables depending only
on the unique model parameter $U$\footnote{If no currents are flowing
  along the string, Lorentz invariance implies that the string tension
  $T$ equals the energy density $U$.}. In Sec.~\ref{sec:smapng}, we
recap the expected CMB temperature anisotropies induced by cosmic
strings, derived from various methods. Particular attention is paid to
small angle CMB maps which preserve all of the projected statistical
information. We then derive the cosmic string signals expected in
various non-Gaussian estimators ranging from the one-point function of
the CMB temperature fluctuations to the bispectrum and trispectrum. We
conclude in Sec.~\ref{sec:conc} and discuss various non-Gaussian
aspects which still have to be explored.

\section{Cosmic strings of various origins}
\label{sec:kinds}

Cosmic strings of cosmological interest can be of various kinds
depending on the microscopic model they stem from. As mentioned in the
introduction, they can either be non-trivial stable, or metastable,
field configurations or more fundamental objects in String
Theory. From a gravitational point of view, they all are however
line-like energy density and pressure distributions. In the following,
we briefly review the different kinds of string having a cosmological
interest and we emphasize their similarities and differences.

\subsection{Abelian vortices}

\begin{figure}
\begin{center}
\includegraphics[width=14cm]{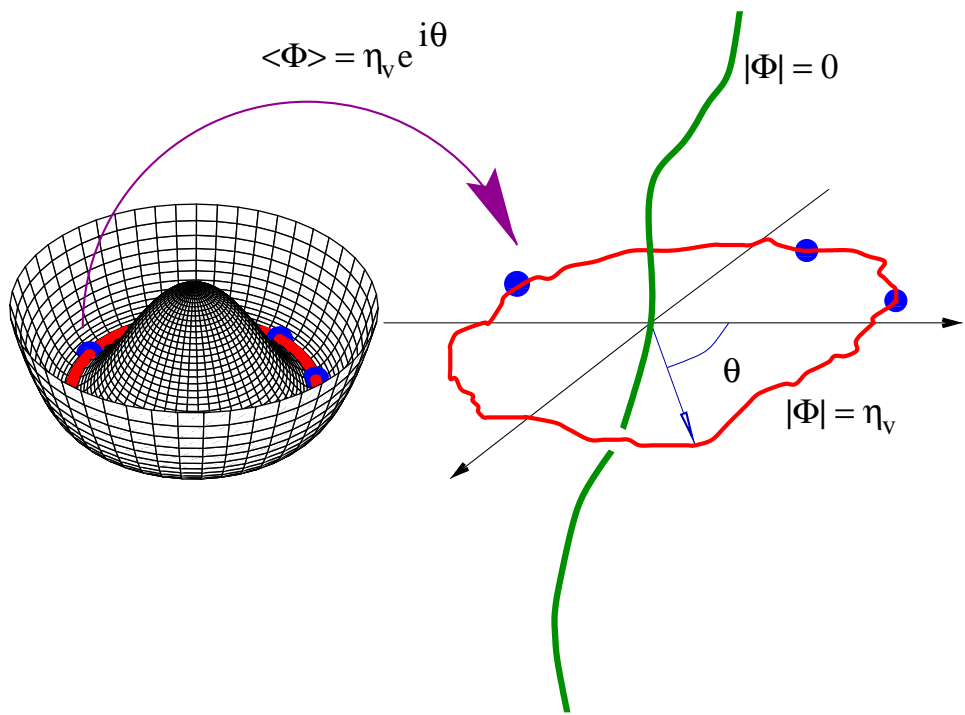}
\caption{The Abelian Higgs potential in the complex plane
  $\left[\Re\left( \Phi \right),\Im\left( \Phi \right) \right]$. The
  non-trivial phase mapping from the internal space to the physical
  space (right) leads to the formation of a cosmic string. The old
  vacuum $|\Phi|=0$ becomes trapped inside the new one $|\Phi|=\etav$.}
\label{fig:higgs}
\end{center}
\end{figure}

The simplest example of cosmic string illustrating the Kibble
mechanism is the Abelian Higgs model. The theory is invariant under a
local gauge group $U(1)$ and the Higgs potential assumes its standard
Mexican hat renormalisable form
\begin{equation}
\label{pothiggs}
V(\Phi) = \frac{\lambda}{8} \left(|\Phi|^2 - \etav^2 \right)^2,
\end{equation}
where $\lambda$ is the self-coupling constant and $\etav$ the vacuum
expectation value of the Higgs field $\Phi$. In Minkowski space, the
Lagrangian reads
\begin{equation}
\label{eq:laghiggs}
\calL_\uh=\frac{1}{2} \left(D_\mu \Phi\right)^\dag \left(D^\mu \Phi
\right) - \frac{1}{4} H_{\mu \nu} H^{\mu \nu} - V(\Phi),
\end{equation}
where $H_{\mu \nu}$ is the field strength tensor associated with the
vector gauge boson $B_\mu$ and
\begin{equation}
D_\mu = \partial_\mu + i g B_\mu.
\end{equation}
At high enough temperature, loop corrections from the thermal bath
restore the $U(1)$ symmetry and the effective potential has an overall
minimum at $|\Phi|=0$~\cite{Kirzhnits:1972, Weinberg:1980}. Starting
from high enough temperature, one therefore expects the $U(1)$
symmetry to be spontaneously broken during the expansion and cooling
of the universe. During the phase transition, the Higgs field reaches
its new vacuum expectation value $\Phi = \etav \ue^{i\theta}$. At each
spacetime location, the phase $\theta(x^\mu)$ will have a given value,
all of them being uncorrelated on distances larger than the typical
correlation length of the phase transition. As pointed by Kibble, this
is at most the horizon size $\horizon \propto t$ although one expects
it to be much smaller~\cite{Kibble:1976, Gill:1994ye, Karra:1997it,
  Bettencourt:1995pj, Kavoussanaki:2000tj}. As a result, there exists
closed paths in space along which $\theta$ varies from $0$ to $2\pi$
(or a multiple of $2\pi$). Such phase configurations necessarily
encompass a point at which $|\Phi|=0$ (see Fig.~\ref{fig:higgs}): the
old vacuum has been trapped into a non-trivial configuration of the
new vacuum, and this prevents its decay. Such a structure is invariant
by translations along the third spatial dimension and is string
shaped.

\begin{figure}
\begin{center}
\includegraphics[width=9cm]{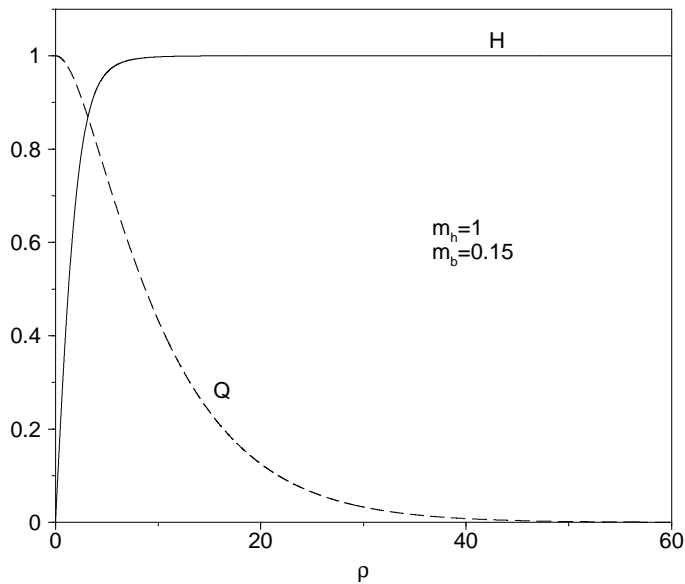}
\caption{String forming field profiles in the Abelian Higgs model with
  unity winding number. The Higgs field $H$ vanishes in the vortex
  core and reaches its vacuum expectation value within distances given
  by the Higgs Compton wavelength $1/m_\uh$. The gauge field $Q$
  condenses inside the vortex over distances given by $1/m_\ub$.}
\label{fig:vortex}
\end{center}
\end{figure}

Solitonic solutions of the field equations describing a static
straight Abelian string can easily be computed under the
Nielsen--Olesen ansatz. The transverse profile of the Higgs and gauge
field are assumed to be~\cite{Nielsen:1987fy}
\begin{equation}
\label{eq:nsansatz}
\Phi = \etav H(\varrho)  \ue^{i n \theta}, \qquad 
B_\mu = \dfrac{Q(\varrho) - n}{g}
\delta_{\mu \theta},
\end{equation}
where $(r,\theta)$ stands for a polar coordinate system aligned along
the string. The dimensionless radial coordinate has been defined by
$\varrho= m_\uh r$ where $m_\uh=\sqrt{\lambda} \etav$ is the mass of
the Higgs boson. The integer $n$ is the ``winding number'' and gives
the number of times the Higgs winds the potential for one rotation
around the string. From Eq.~(\ref{eq:laghiggs}), the dimensionless
equations of motion read
\begin{equation}
\frac{\ud^2 H}{\ud \varrho^2}+\frac{1}{\varrho} \frac{\ud H}{\ud
\varrho}  =  \frac{H Q^2}{\varrho^2}+\frac{1}{2}H(H^2-1), \qquad
\frac{\ud^2 Q}{\ud \varrho^2} -\frac{1}{\varrho}\ \frac{\ud Q}{\ud
\varrho}  =  \frac{m_\ub^2}{m_\uh^2}H^2 Q,
\end{equation}
where $m_\ub=g \etav$ is the mass of the vector gauge boson. In
Fig.~\ref{fig:vortex}, we have represented the string solution to
these equations in Minkowski spacetime~\cite{Peter:1992dw,
  Ringeval:2000kz}. The boundary conditions are such that the Higgs
field vanishes at the center of the string to reach its vacuum
expectation value (vev) asymptotically. This typically happens after a
length scale given by its Compton wavelength $1/m_\uh$. Similarly, the
gauge field boundary conditions are such that it has vanishing
derivative in the core and remains finite far from the string. As
shown in Fig.~\ref{fig:vortex}, it actually condenses inside the
string with a spatial extension roughly equals to $1/m_\ub$.

The energy content of such a string is given by the stress tensor
stemming from the Lagrangian of Eq.~(\ref{eq:laghiggs}). Along the
string worldsheet,
\begin{equation}
\label{eq:tmunuvortex}
T^{tt} = - T^{zz} = \frac{\lambda \etav^4}{2}
\left[ \left(\partial_\varrho H
  \right)^2 + \frac{Q^2 H^2}{\varrho^2} + \frac{(H^2-1)^2}{4} +
  \frac{\lambda}{g^2} \frac{\left( \partial_\varrho Q
    \right)^2}{\varrho^2} \right],
\end{equation}
which are the only two components which do not vanish after an
integration over the transverse coordinates. Integrating the temporal
part gives the string energy per unit length $U$, whereas the
longitudinal component gives $-T$. One finally gets
\begin{equation}
U = T = C\!\left(\lambda/g^2 \right) \, \etav^2,
\end{equation}
where $C(\lambda/g^2)$ is an order unity function at fixed winding
number. Increasing the winding number centrifuges the energy density
around the core such that $U$ is changed in a more complex
way~\cite{Vilenkin:2000}. This immediately shows that cosmic strings
generically carry an energy density and tension of the order of the
symmetry breaking energy scale $U \simeq \etav^2$. Notice that along
the string direction the pressure $P_z=-T=-U$ is negative, and we are
in presence of a ``cosmological constant wire'', as one may expect
from a Lorentz invariant vacuum object. Consequently, the trace of the
stress tensor $\eta_{\mu \nu} T^{\mu \nu}$ vanishes and cosmic strings
do not induce any Newtonian gravitational potential. Together with the
so-called cosmological scaling behaviour (see below), this is the mere
reason why they remain cosmologically acceptable. They do however
induce dynamical gravitational effects, the metric far from the string
core being Minkowski with a missing angle~\cite{PhysRevD.23.852} (see
Sec.~\ref{sec:smapng}).

The Abelian string model is intensively used in the literature to
explore the string forming phase transition and string
interactions~\cite{Bettencourt:1994kf, Antunes:1997pm}. From the
Kibble's argument, one expects the phase of the Higgs field to be
random and the resulting string path should be a self-avoiding random
walk with a given correlation
length~\cite{Vachaspati:1984}. Performing lattice simulations allows
to probe in more details the string forming mechanism and gives a more
accurate picture of a cosmic string network just after its
formation~\cite{Hindmarsh:2001vp, Donaire:2005qm, Rajantie:2008bc,
  Sakellariadou:2008ay}. Abelian Higgs simulations are also used to
compute the cosmological evolution of such a
network~\cite{Vincent:1998, Moore:2002, Hindmarsh:2008dw} (see
Sec.~\ref{sec:evol}).

\subsection{Other flux tubes}

\subsubsection*{Global string.} The Abelian string provides an
explicit example of the formation of line-like topological defects by the
spontaneous breakdown of a gauged symmetry. Breaking a $U(1)$ global
symmetry can also produce topological defects, the so-called global
strings. However, in the absence of gauge fields, one can show that
global cosmic strings exhibit long range interactions and Goldstone
radiation~\cite{PhysRevD.32.3172}. Their dynamics can however mimic
local strings and being cosmologically acceptable in some
regime~\cite{Durrer:1998rw, Yamaguchi:1999yp, Yamaguchi:1999dy}.

\subsubsection*{Non-Abelian string.} If the broken symmetry group $G$
is non-Abelian, the cosmic strings formed during the phase transition
exhibit new properties compared to the $U(1)$
kind~\cite{Vilenkin:2000}. In particular, the mapping of the Higgs
field to the real space can be made along different broken generators
$\Phi_a = \etav \exp(i T_a \theta)$. This implies that different type
of non-Abelian strings may be formed and will interact with each
others according to their respective windings. The classic example
being the appearance of a $[T_a,T_b]$-string from the crossing between
a $T_a$-string and $T_b$-string~\cite{PhysRevD.57.3317}. In the
cosmological framework, new strings can potentially be formed at each
interaction leading to a frustrated intricated
configuration~\cite{Vilenkin:1984rt, Dvali:1993qp, Spergel:1996ai,
  Bucher:1998mh}. Such an outcome depends on the underlying
non-Abelian gauge group and Abelian string-like evolution can also be
recovered, as for instance in the $U(N)$
models~\cite{Hashimoto:2005hi, Eto:2006db, Eto:2006pg,Shifman:2007ce,
  2009PhRvL.103k5301K}.

\subsubsection*{Semi-local string.} String-shaped energy density
distribution can also appear even if the vacuum manifold is simply
connected. The non-trivial topology argument is indeed only a
sufficient condition of defect appearance. The electroweak symmetry
breaking scheme enters this class, although the first homotopy group
is trivial, semi-local strings can be formed~\cite{PhysRevD.44.3067,
  Achucarro:1998ux}. A simple description of these strings can be
obtained by replacing the Higgs field in the Abelian model by a
doublet in a $SU(2)$ global
representation~\cite{Achucarro:1999it}. These flux tube configurations
are stabilised because they can be energetically favoured for some
values of the model parameters, typically for $m_\ub >
m_\uh$~\cite{Hindmarsh:1991jq}. Let us notice that the currently
measured electroweak model parameters do not support stable
vortices~\cite{Achucarro:1999it}. Similar vortices could also be
formed during a chiral symmetry breaking phase
transition~\cite{Zhang:1997is, Balachandran:2001qn, Buckley:2002ur,
  Brandenberger:2002qh, Nitta:2007dp, Eto:2009wu} or within strong
interaction models~\cite{Nakano:2008dc, Eto:2009tr}.

\subsubsection*{K- and DBI-string.}
These are another extensions of the Abelian Higgs string for which the
scalar and gauge field kinetic terms are non-canonical, or of the
Dirac--Born--Infeld form~\cite{Babichev:2006cy, Babichev:2007tn,
  Sarangi:2007mj, Babichev:2008qv}. These strings essentially differ
from their Abelian counterparts when the gradient terms are
non-vanishing, i.e. in the core.

\subsubsection*{Current carrying string.}
In minimal extensions of the Abelian Higgs model, one may couple
extra-scalar fields to the string forming Higgs field. As shown by
Witten, this can lead to the condensation of the extra-scalar over the
string core~\cite{Witten:1984eb}. The resulting string is carrying a
current that breaks the longitudinal Lorentz invariance: $U$ and $T$
are no longer degenerate and the string dynamics is
affected~\cite{Carter:1989dp, Peter:1992dw, Carter:2000wv}. One of the
most important consequence of these currents is the potential
appearance of centrifugally supported loops. If stable, these
so-called vortons could efficiently populate the universe and avoiding
the overclosure gives strong constraints on the cosmic string energy
scale~\cite{Davis:1988ip, Brandenberger:1996zp}. A similar mechanism
works for the fermionic fields which are Yukawa coupled to the string
forming Higgs field. They generically produce currents along the
string with a discrete mass spectrum, in a way similar to the photon
propagation in waveguides~\cite{Ringeval:2001xd}. Unless the massive
propagation modes are not excited, the resulting loops are however
expected to be unstable~\cite{Carter:1993wu, Martin:1994qh}.

\subsection{Cosmic superstrings}

Cosmic superstrings are fundamental line-shaped objects that can be
formed at the end of brane-inflation (see Refs.~\cite{Copeland:2009ga,
  Sakellariadou:2008ie, Sakellariadou:2009ev, Polchinski:2004ia,
  Davis:2008dj} for reviews). The idea that fundamental quantum
strings can be stretched to cosmological distances has been mentioned
by Witten~\cite{Witten:1985fp}. If stable, one would expect
fundamental strings to be at an energy scale close the String Theory
scale, i.e. close to the Planck mass, and this is trivially ruled out
by observations. In addition, current CMB constraints tell us that the
energy scale of inflation is at most the GUT
scale~\cite{Lorenz:2008je}, implying that strings formed at an higher
energy would have been diluted anyway. The situation changed with the
discovery that inflation within String Theory could be a geometrical
phenomena induced by the motion of a brane moving in a warped throat,
somewhere in the compact manifold of the
extra-dimensions~\cite{Dvali:1998pa}. In the KKLMMT
model~\cite{Kachru:2003sx, Giddings:2001yu}, the inflaton is a scalar
degree of freedom associated with the position $r$ of a D3-brane in a
warped throat. Within a ten-dimensional super-gravity ansatz for the
metric, in the type IIB String Theory,
\begin{equation}
\ud s^2 = \dfrac{1}{\sqrt{h(r)}} g_{\mu \nu} \ud x^\mu \ud x^\nu +
\sqrt{h(r)} \left(\ud r^2 + r^2 \ud_{_5} s^2\right),
\end{equation}
the throat is described by the warping function $h(r)$ (explicitly, it
can be the Klebanov--Strassler conifold~\cite{Klebanov:2000hb}). In
this system, accelerated expansion of our universe comes from the
interaction of this brane with an anti D3-brane sitting at the bottom
of the throat $r_0$. Current CMB data suggest that inflation
preferentially ends by violation of the slow-roll conditions whereas
the system continues to evolve till the two branes
collide~\cite{Lorenz:2007ze}. The brane interactions at that stage
require String Theory calculations and are expected to trigger a
reheating era accompanied by a copious production of various D1-branes
and fundamental F-strings~\cite{Burgess:2001fx, Sarangi:2002yt,
  Dvali:2003zj, Jones:2003da}. Since the brane annihilation takes
place at the bottom of the throat, due to the warped metric, the
cosmic superstring tensions measured by an exterior four-dimensional
observer are redshifted by a factor $h^{1/2}(r_0)$. The resulting
effect is to significantly lower the string tension down to acceptable
values. In fact, the stability of the produced F-strings and D-strings
require additional constraints on the model parameters and the
spectrum of superstring tensions depends on the underlying
scenario~\cite{Jones:2003da}. For instance, in the KKLMMT model, one
expects $10^{-10}<GU<10^{-7}$~\cite{Firouzjahi:2005dh}.

Cosmic superstrings differ from the Abelian strings in various
aspects. In addition to the coexistence of two different types, they
can form bound states of $p$ F-strings and $q$ D-strings. The tension
of these $(p,q)$-strings depends on $p$, $q$, the binding energy but
also on their configuration in the throat~\cite{Gubser:2004qj,
  Firouzjahi:2006vp}. In fact, many of $(p,q)$-string properties mimic
the non-Abelian type of topological vortex, as the existence of bound
states and Y-junctions~\cite{Jackson:2006qc, Copeland:2006if,
  Rivers:2008je, Bevis:2009az}. Such similarities have actually been
used to probe the properties of the cosmic superstrings through the
more tractable framework of field theory~\cite{Saffin:2005cs,
  Cho:2005wp, Hindmarsh:2006qn, Rajantie:2007hp}.

\subsection{Infinitely thin strings}

These are the one-dimensional version of the relativistic point
particles. Following Carter macroscopic covariant
approach~\cite{Carter:1989dp, Carter:2000wv, Carter:1989xk,
  Carter:1992vb}, string events can be localised in the
four-dimensional spacetime by the so-called embedding functions
$x^\mu=X^\mu(\xi^a)$, where $\xi^0$ and $\xi^1$ are a timelike and
spacelike internal coordinate of the string worldsheet. Denoting by
$g$ the four-dimensional metric tensor, one can define the
two-dimensional induced metric
\begin{equation}
  \gamma_{ab} = g_{\mu \nu} \dfrac{\partial X^\mu}{\partial \xi^a} \dfrac{\partial
    X^\nu}{\partial \xi^b}\,,
\end{equation}
such that the infinitesimal interval between two events reduces to
$\ud s^2 = \gamma_{ab} \ud \xi^a \ud \xi^b$. From its inverse, on can
define the first fundamental tensor
\begin{equation}
  q^{\mu \nu} = \gamma^{a b} \dfrac{\partial X^\mu}{\partial \xi^a}
  \dfrac{\partial X^\nu}{\partial \xi^b},
\end{equation}
which is nothing but a projector over the string
worldsheet. Similarly, $\perp^\mu_{\phantom{\mu}\nu}\equiv
g^\mu_{\phantom{\mu}\nu} - q^\mu_{\phantom{\mu}\nu}$ is an orthogonal
projector and they verify
\begin{eqnarray}
  q^\mu_{\phantom{\mu}\rho} q^\rho_{\phantom{\rho} \nu} 
= q^\mu_{\phantom{\mu} \nu}, \qquad
  \perp^\mu_{\phantom{\mu} \rho} \perp^\rho_{\phantom{\rho} \nu} 
= \perp^\mu_{\phantom{\mu} \nu}, \qquad
  q^\mu_{\phantom{\mu} \rho} \perp^\rho_{\phantom{\rho} \nu} =0.
\end{eqnarray}
Variations of the first fundamental form are encoded in the second
fundamental tensor
\begin{equation}
K_{\mu \nu}^{\phantom{\mu \nu} \rho} \equiv q^\alpha_{\phantom{\alpha}\nu}
  \nablab_\mu \, q^\rho_{\phantom{\rho} \alpha},
\end{equation}
where a bar quantity stands for the projection of its four-dimensional
analogue over the worldsheet, i.e. $\nablab_\mu \equiv
q^\alpha_{\phantom{\nu}\mu} \nabla_\alpha$. Integrability imposes
$K_{[\mu \nu]}^{\phantom{[\mu\nu]}\rho}=0$ and, by construction, the
second fundamental form is, respectively, tangent and orthogonal to
the worldsheet on its first and last indices. As a result, contracting
the first two tangential indices gives a purely orthogonal vector
which measures the string extrinsic curvature~\cite{Carter:1992vb}
\begin{equation}
\label{eq:excurv}
  K^\rho \equiv K^{\mu\phantom{\mu}\rho}_{\phantom{\mu}\mu} =
  \nablab_\mu  q^{\mu \rho}.
\end{equation}

The energy content of a spacetime two-dimensional surface can be
characterised by its internal stress energy tensor. Similarly to the
cosmological perfect fluid, one may consider a string whose
stress-energy tensor is diagonal in a preferred basis. Positivity of
the energy conditions ensures that the timelike eigenvalue $U>0$,
while the spacelike eigenvalue $T$ should verify $|T| \le
U$~\cite{Wald:553440}. In this frame, $U$ represents the energy per
unit length of the string and $T$ the string tension. Denoting by
$u^\mu$ and $v^\nu$ the respective timelike and spacelike orthonormal
eigenvectors, one has
\begin{equation}
\label{eq:stress}
  \Tb^{\mu \nu} = U u^\mu u^\nu - T v^\mu v^\nu = (U-T) u^\mu u^\nu - T
  q^{\mu \nu},
\end{equation}
where
\begin{equation}
  u^\alpha u_\alpha = -1, \qquad v^\alpha v_\alpha = 1, \qquad u^\alpha
  v_\alpha = 0, \qquad q^{\mu \nu} = -u^\mu u^\nu + v^\mu v^\nu.
\end{equation}
In the absence of external forces, reparametrisation invariance of the
string worldsheet ensures the stress-energy pseudo-conservation from
Noether's theorem~\cite{PhysRevD.50.682}
\begin{equation}
\label{eq:divstress}
\nablab_\rho \Tb^{\rho \sigma} = 0.
\end{equation}
As for a cosmological fluid, these equations are not sufficient to
close the equations of motion for the string. One has to supplement
them by an equation of state of the fluid under scrutiny. The simplest
case is the so-called barotropic model for which the equation of state
is the relation $U(T)$. One can then introduce the two Legendre
conjugated parameters
\begin{equation}
\ln \nub = \int \dfrac{\ud U}{U-T}\,, \qquad \ln \mub = \int \dfrac{\ud T}{T-U}\,,
\end{equation}
such that $U-T=\mub \nub$. Clearly, $\nub$ plays the role of a number
density and its Legendre conjugated parameter $\mub$ will therefore be
a chemical potential, i.e. an effective mass carried per unit number
density. Defining their respective worldsheet current density by
\begin{equation}
\mub^\rho \equiv \mub u^\rho, \qquad \nub^\rho \equiv \nub u^\rho,
\end{equation}
one can rewrite Eq.~(\ref{eq:divstress}) as
\begin{equation}
\label{eq:baromotion}
  \nablab_\rho \Tb^{\rho}_{\phantom{\rho}\sigma} = 
  \mub_\sigma \nablab_\rho \nu^\rho + \nub^\rho \nablab_\rho \mub_\sigma
  - \nub^\rho \nablab_\sigma \mub_\rho -T K_\sigma = 0.
\end{equation}
Contracting Eq.~(\ref{eq:baromotion}) with $\nub^\sigma$ ensures the
current conservation along the string
\begin{equation}
\label{eq:current}
\nablab_\rho \nub^\rho = 0,
\end{equation}
while its projection onto the worldsheet gives the momentum transport
law
\begin{equation}
\label{eq:kelvin}
q^\sigma_{\phantom{\sigma} \alpha} u^\rho \nablab_{\left[\rho
\right.} \mub_{\left. \sigma \right]} = 0.
\end{equation}
Finally, the orthogonal projection of Eq.~(\ref{eq:baromotion})
reduces to
\begin{equation}
\label{eq:extmotion}
K^\rho = \perp^\rho_{\phantom{\rho}\sigma}
\dfrac{\nub^\alpha \nablab_\alpha \mub^\sigma}{T}
= \perp^\rho_{\phantom{\rho}\sigma} \left(\dfrac{U}{T}-1\right) \dot{u}^\sigma,
\end{equation}
where the string acceleration $\dot{u}^\sigma$ stands for
\begin{equation}
  \dot{u}^\sigma \equiv u^\alpha \nablab_\alpha
  u^\sigma.
\end{equation}

As should be clear from Eq.~(\ref{eq:stress}), the barotropic equation
of state breaks Lorentz invariance along the string for $U \neq T$. In
fact, it describes a wide class of elastic string
models~\cite{PhysRevD.41.3869, Carter:1994zs, Carter:2003fb}, and as
suggested by Eq.~(\ref{eq:current}), the scalar current carrying
cosmic strings~\cite{Carter:1999hx}. Conversely, imposing Lorentz
invariance along the worldsheet reduces the equation of state to the
trivial form $U=T$ [see Eq.~(\ref{eq:stress})], which is also the
relation found for the Abelian Higgs string. This infinitely thin
string is the Nambu--Goto (NG) string and does not possess any
internal structure~\cite{Goto:1971ce}. The associated equations of
motion are purely geometrical and do not depend on $U$. From
Eq.~(\ref{eq:extmotion}), they reduce to the vanishing of the
extrinsic curvature vector, i.e.
\begin{equation}
K^\rho = 0, 
\end{equation}
which can be rewritten in a coordinate dependant way by using
Eq.~(\ref{eq:excurv})
\begin{equation}
\label{eq:excurvfull}
K^\mu  = \frac{1}{\sqrt{-\gamma}}\partial_a \left(\sqrt{-\gamma}
\gamma^{a b} \partial_b X^\mu \right) + \Gamma^\mu_{\nu \rho}
\gamma^{a b} \partial_a X^\nu \partial_b X^\rho.
\end{equation}
The connections $\Gamma^\mu_{\nu\rho}$ are for the background
spacetime of metric $g_{\mu \nu}$ while $\gamma$ is the determinant of
the induced metric. These equations can also be recovered from the
usual NG action with an explicit coordinate
system~\cite{Vilenkin:2000}
\begin{equation}
\label{eq:ngaction}
  S = - U \int \ud^2 \xi \sqrt{-\gamma}\,.
\end{equation}

\section{Cosmological evolution of Nambu--Goto strings}
\label{sec:evol}

The previous section shows that the equations of motion of an isolated
string depend on the underlying microscopic model. The type of string
is more determinant when two strings interact: cosmic superstrings may
form bound states, while non-Abelian vortices may weave new vortices
from each of their interaction points. Understanding the cosmological
evolution of a string network requires one to solve both the local
equations of motion for each string and the outcome of their
interactions when they meet. Moreover, the evolution of a system of
strings starts from an initial configuration which should describe the
network configuration just after its formation. Numerical simulations
have been used to overcome some of the above-mentioned difficulties
and, up to now, Friedmann--Lema\^{\i}tre--Robertson--Walker (FLRW)
network simulations have only been performed with Nambu--Goto strings,
Abelian Higgs strings and semi-local strings~\cite{Bennett:1989,
  Albrecht:1989, Bennett:1990, Allen:1990, Vincent:1998, Moore:2002,
  Ringeval:2005kr, Martins:2005es, Olum:2006ix, Hindmarsh:2008dw,
  2008JCAP.07.010U}, up to some variations~\cite{Yamaguchi:1999yp,
  Yamaguchi:1999dy, Hindmarsh:2006qn, Urrestilla:2007yw}. As a result,
extrapolating the following results to other types of string should be
made with caution. On the bright side, Eq.~(\ref{eq:extmotion})
suggests that as long as the string acceleration remains small
compared to $T/(U-T)$, one expects the equations of motion of the
string to be close to the NG case (up to the eventual vortons
appearance). In the following, we describe the results obtained for NG
strings. Some differences exist with the results obtained in the
Abelian Higgs string simulations.

Before entering into details, let us summarize two fundamental
properties these simulations have revealed. The first is that a cosmic
string network avoids cosmological domination by evacuating most of
its excess energy through some complex mechanisms, which typically
result in transferring energy between the horizon-sized distances and
the smaller length scales. For NG simulations, this is the formation
of cosmic string loops whereas in Abelian Higgs simulation boson
radiation is involved. The second property is that the influence of
the initial conditions is expected to disappear on the length scales
of astrophysical interests. A network of cosmic strings relaxes
towards a cosmological attractor which depends only on the expansion
rate: this is the so-called scaling regime.

\subsection{Dynamics}

The equations of motion for NG strings are the vanishing of the
extrinsic curvature vector $K^\mu = 0$. In a flat FLRW background,
\begin{equation}
  \ud s^2 = a^2(\eta) \left(-\ud \eta^2 + \delta_{ij} \ud x^i \ud x^j \right),
\end{equation}
Eq.~(\ref{eq:excurvfull}) can be simplified with the transverse gauge fixing
conditions
\begin{equation}
\label{eq:transgauge}
  g_{\mu\nu} \dfrac{\partial X^\mu}{\partial \tau} \dfrac{\partial
    X^\nu}{\partial \sigma} = 0,
\end{equation}
with the notation $\tau = \xi^0$ and $\sigma = \xi^1$ for the timelike
and spacelike string coordinates. Such a choice of coordinates
reflects the property that a NG string is Lorentz invariant along the
worldsheet: there is no physical longitudinal component of the string
velocity. In this gauge, the equations of motion read
\begin{eqnarray}
\label{eq:ngtrans}
\Xdd^\mu   + \left( \frac{\dot\varepsilon}{\varepsilon}  +  \frac{2}{a} \frac{\ud 
    a}{\ud X^0} \Xd^0 \right) \Xd^\mu & - &
\frac{1}{\varepsilon} \left(\frac{\Xp^\mu}{\varepsilon} \right)'  
 \nonumber \\ & - &  \frac{2}{a} \frac{\ud a} {\ud X^0}
\frac{\Xp^0}{\varepsilon} \frac{\Xp^\mu}{\varepsilon} 
+  \delta^\mu_0 \frac{2}{a} \frac{\ud a}{\ud X^0}\Xd^2 = 0,
\end{eqnarray}
where a ``dot'' and a ``prime'' stand respectively for differentiation
with respect to $\tau$ and $\sigma$. We have also defined the quantity
\begin{equation}
\label{eq:epsdef}
\varepsilon \equiv \sqrt{-\dfrac{\Xp^2}{\Xd^2}}\,.
\end{equation}
The conditions in Eq.~(\ref{eq:transgauge}) do not completely fix the
coordinate degrees of freedom and one can supplement them with the
so-called temporal gauge fixing which identifies the timelike
coordinate with the background time at the string event: $\tau =
X^0=\eta$. In the transverse temporal gauge Eq.~(\ref{eq:transgauge})
reads $\vecXd \cdot \vecXp = 0$, while Eq.~(\ref{eq:ngtrans})
simplifies to
\begin{eqnarray}
\label{eq:ngtranstemp}
  \vecXdd + 2\calH \left(1- \vecXd^2 \right) - \dfrac{1}{\varepsilon} \left(
    \dfrac{\vecXp}{\varepsilon} \right)' = 0, \qquad \dot{\varepsilon} + 2
  \calH \varepsilon \vecXd^2 = 0, 
\end{eqnarray}
with
\begin{equation}
\varepsilon=\sqrt{\dfrac{\vecXp^2}{1-\vecXd^2}}\,,
\end{equation}
and $\calH$ is the conformal Hubble parameter. The vector symbols
being understood as three-dimensional spatial vectors. Numerically, it
is much more convenient to solve an equivalent set of equations found
by Bennett and Bouchet~\cite{Bennett:1990}. Defining the new vectors
$\bp$ and $\bq$ as
\begin{equation}
\label{eq:bbkinks}
\bp(\tau,u) \equiv \dfrac{\vecXp}{\varepsilon} - \vecXd, \qquad \bq(\tau,v)
\equiv \dfrac{\vecXp}{\varepsilon} + \vecXd,
\end{equation}
evaluated at the new coordinates $u=\int \varepsilon \ud \sigma - \tau$
and $v=\int \varepsilon \ud \sigma + \tau$, the equations of motion
(\ref{eq:ngtranstemp}) can be recast into
\begin{eqnarray}
\label{eq:pqevol}
  \dfrac{\partial \bp}{\partial \tau} &=& -\calH \left[\bq -
    \bp\left(\bp\cdot\bq \right) \right], \qquad
  \dfrac{\partial \bq}{\partial \tau} = -\calH \left[\bp -
    \bq\left(\bp\cdot\bq \right) \right], \\ 
\label{eq:epsevol}
\dfrac{\dot{\varepsilon}}{\varepsilon} & = &  -\calH \left(1 - \bp \cdot \bq \right).
\end{eqnarray}
As an illustrative example, these equations have an exact solution in
Minkowski space. Taking $\calH=0$, one immediately gets
$\varepsilon=1$ (up to a normalisation constant), $\bp(u)$ and
$\bq(v)$ are constant over the characteristics $u=\sigma-\tau$ and
$v=\sigma+\tau$. Inverting Eq.~(\ref{eq:bbkinks}) gives
\begin{equation}
  \vecXp(\tau,\sigma) = \dfrac{1}{2} \left[\bp(\sigma+\tau) + \bq(\sigma -\tau)\right],
\end{equation}
which describes the propagation of left and right moving string
deformations at the speed of light. In the FLRW background, these
modes are no longer free moving, but interact through the Hubble term
[see Eq.~(\ref{eq:pqevol})]. Solving these equations gives the
$X^\mu(\tau,\sigma)$ for each strings but does not predict what
happens when two strings collides.

\subsection{Collisions}
\begin{figure}
  \begin{center}
    \includegraphics[height=3cm]{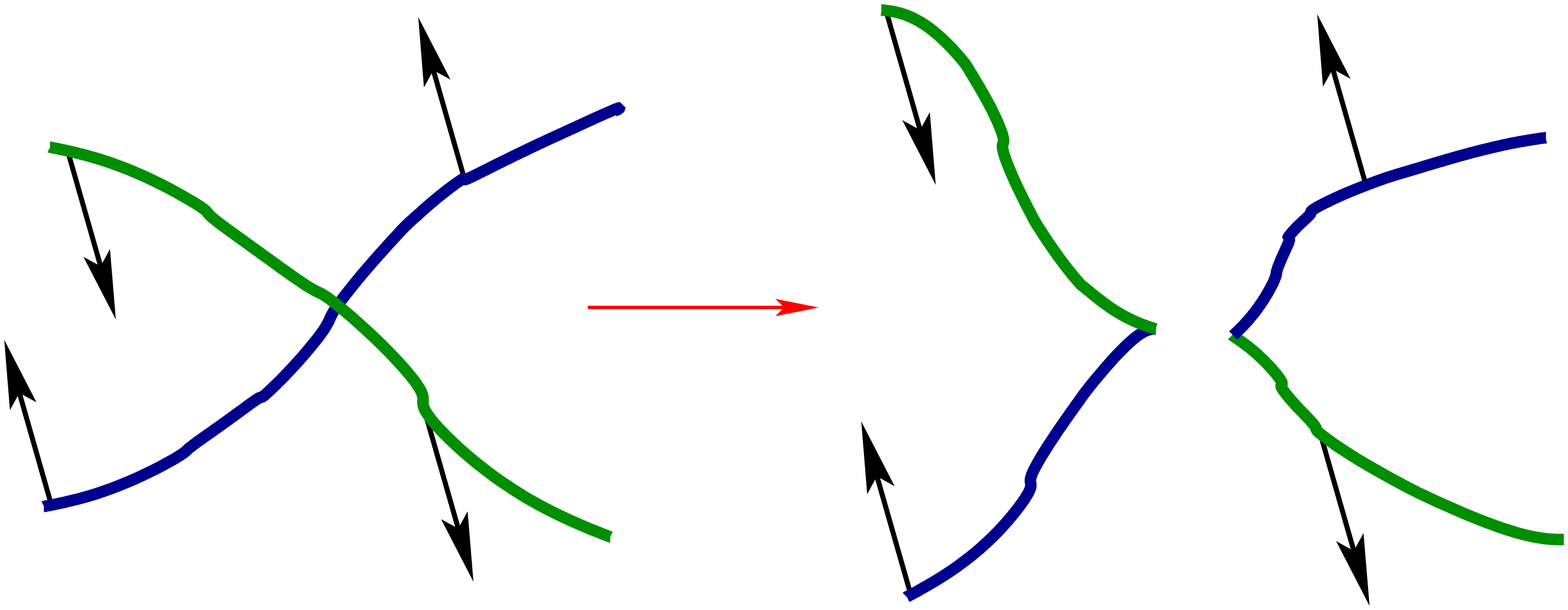}
    \hspace{1cm}
    \includegraphics[height=3cm]{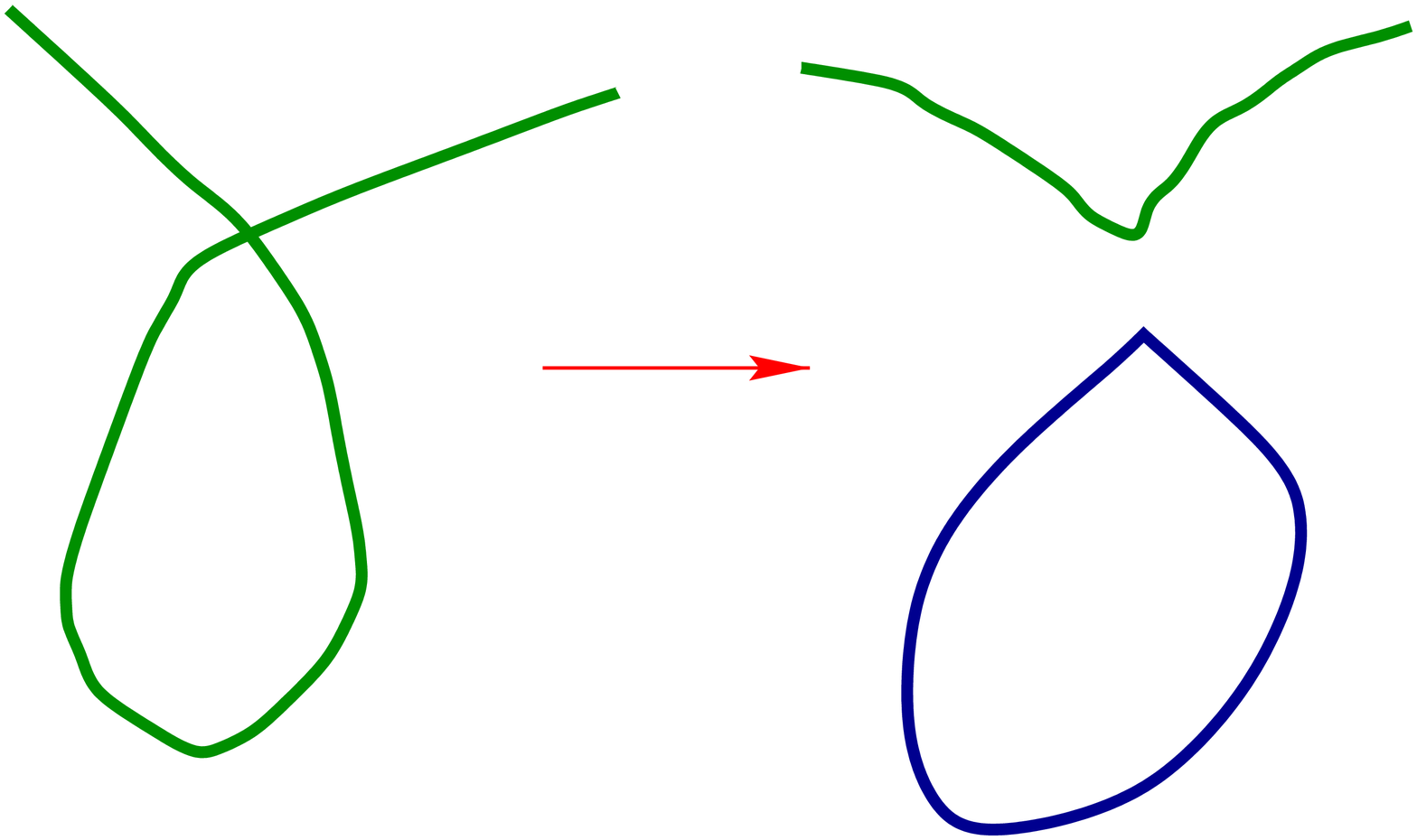}
    \caption{Intercommuting strings exchanging their partners
      (left). On the right, the same mechanism produces a loop from a
      self-intercommutation. The arrows represent velocity vectors.}
    \label{fig:intercom}
  \end{center}
\end{figure}
\begin{figure}
  \begin{center}
    \includegraphics[width=5.0cm,height=5.0cm]{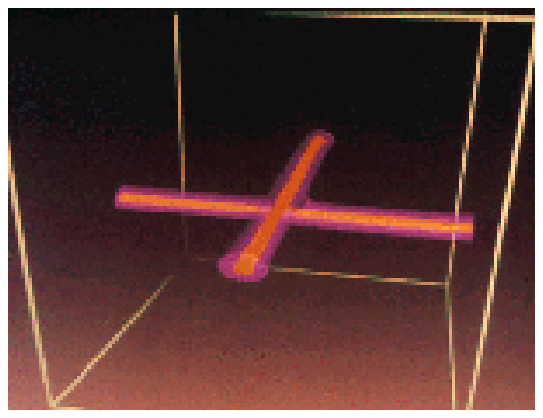}
    \hspace{1.5mm}
    \includegraphics[width=5.0cm,height=5.0cm]{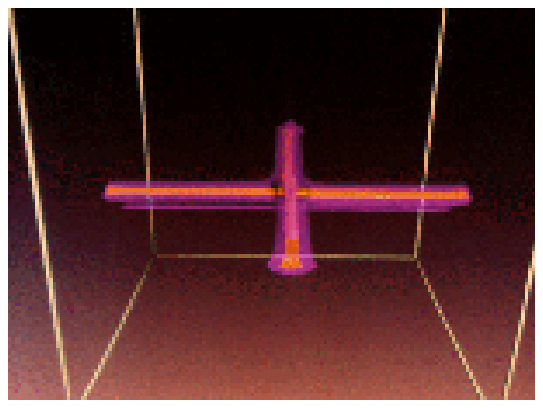}
    \hspace{1.5mm}
    \includegraphics[width=5.0cm,height=5.0cm]{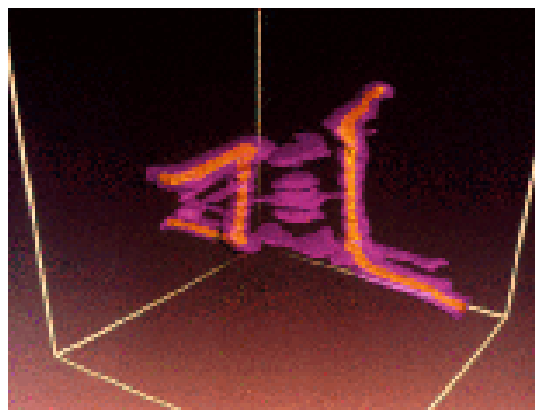}
    \hspace{1.5mm}
    \includegraphics[width=5.0cm,height=5.0cm]{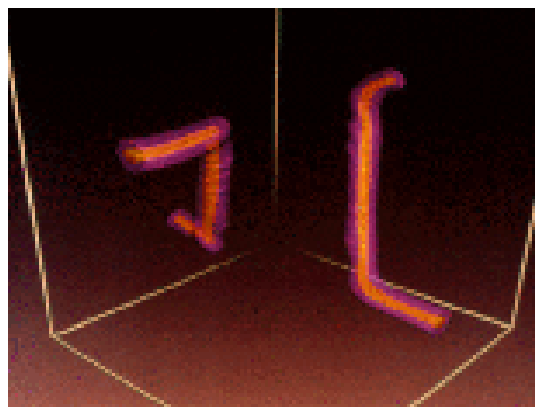}
    \caption{Numerical simulations of the intersection between two
      Abelian Higgs strings~\cite{Matzner:1988}. The Higgs and gauge
      field profiles are represented as the inner orange and outer
      pink tubes. Generically, Abelian Higgs strings with
      $m_\uh>m_\ub$ intercommute.}
    \label{fig:vortexcross}
  \end{center}
\end{figure}

In the infinitely thin approach, the outcome of a NG string
intersection event cannot be predicted. Stress tensor conservation
equations require either that the two strings pass through each
others, or they intercommute as sketched in
Fig.~\ref{fig:intercom}. The outcome of a string collision process can
only be addressed within the framework of a microscopic
model. Numerical simulations of interactions have been performed for a
variety of models, and in particular for the Abelian Higgs string as
represented in Fig.~\ref{fig:vortexcross}. In this case, unless the
relative string velocity is close to unity~\cite{Hanany:2005bc,
  Achucarro:2006es}, or the strings are almost parallel, string
intercommutation generically
occurs~\cite{Shellard:1987bv,Matzner:1988}. Let us notice that for
type I Abelian strings (i.e. having $m_\ub>m_\uh$), bound states of
the two strings can also be formed at low
velocity~\cite{Bettencourt:1994kc, Bettencourt:1996qe, Salmi:2007ah}.

The situation is not necessarily the same for the other types of
string. As already mentioned, non-Abelian strings can weave new
strings from their intersection points while current carrying cosmic
strings intercommute for bosonic carriers~\cite{Laguna:1990}. In the
case of cosmic superstrings, it has been shown that they intercommute
with a probability depending on the fundamental string coupling, a
quantity which can be significantly smaller than
unity~\cite{Jackson:2004zg}. In the case of $(p,q)$-string collisions,
Y-junctions can be formed under some kinematic
constraints~\cite{Copeland:2006if, Copeland:2006eh, Salmi:2007ah,
  Copeland:2007nv, Bevis:2008hg, Binetruy:2010bq}. Concerning NG
simulations, string collisions are actually implemented through a
phenomenological probability $\Pe$ of intercommutation at each
intersection event.

\subsection{Initial conditions}

\begin{figure}
  \begin{center}
    \includegraphics[width=14cm]{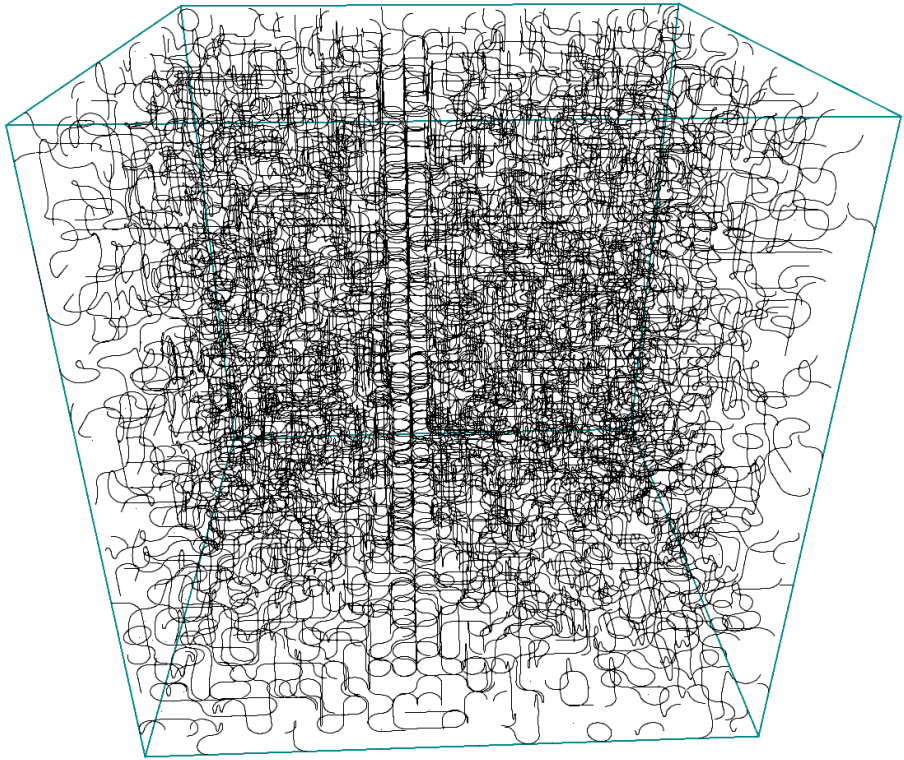}
    \caption{Initial string network configuration from the
      Vachaspati--Vilenkin algorithm. The box is a small lattice of
      $25 \corrini$ for illustration purpose only.}
    \label{fig:vvini}
  \end{center}
\end{figure}

Solving the cosmological evolution of a NG string network amounts to
solving Eqs.~(\ref{eq:pqevol}) and (\ref{eq:epsevol}) along each
string, finding all of their intersection points and implementing an
intercommutation, or not, with the probability $\Pe$. The network
evolution is now uniquely determined once the initial conditions are
specified. The simplest way to set initial conditions is through the
Vachaspati--Vilenkin (VV)
algorithm~\cite{Vachaspati:1984dz}. Motivated by the Kibble mechanism,
one assumes a $U(1)$ Higgs field to be uncorrelated above a given
correlation length $\corrini$. A cosmic string will cross a given
plane if one can find a closed loop along which its phase runs from
$0$ to a multiple of $2\pi$. On a discrete three-dimensional lattice,
of $\corrini$-spacing, it is sufficient to approximate $U(1)$ by $Z_3$
and randomly choose the phase at each corner from three values $0$,
$2\pi/3$ and $4\pi/3$ to decide if a string crosses the associated
face. Other symmetry breaking schemes and lattice can be approximated
in a similar way~\cite{Kibble:1985xe, Yates:1995gj, Borrill:1995gu,
  Scherrer:1997sq, Scherrer:1997dt}. In Fig.~\ref{fig:vvini}, we have
shown the initial string network configuration obtained from the VV
algorithm. The string paths have been smoothed by replacing the right
angles by circles of radius $\corrini$.
The initial network configuration obviously depends on the physical
parameter $\corrini$, the network correlation length. In FLRW
spacetime, there is however another physical parameter which has to be
specified: the distance to the horizon $\horizon$. From those, the
initial string energy density is now uniquely determined. A random
transverse velocity field can also be added along each string since
one does not expect the strings to be initially at rest in any
realistic setup. At this point, let us mention that the numerical
implementation of the VV initial conditions introduce two additional
purely numerical parameters: the size of the periodic box which
contains the simulation, usually normalised to unity in comoving
coordinates, and the discretisation step required to represent a
string, usually given by $\Nppcl$, the number of points per
correlation length.

\subsection{Cosmological scaling}

\subsubsection{Long strings.}
\begin{figure}
  \begin{center}
    \includegraphics[width=14cm]{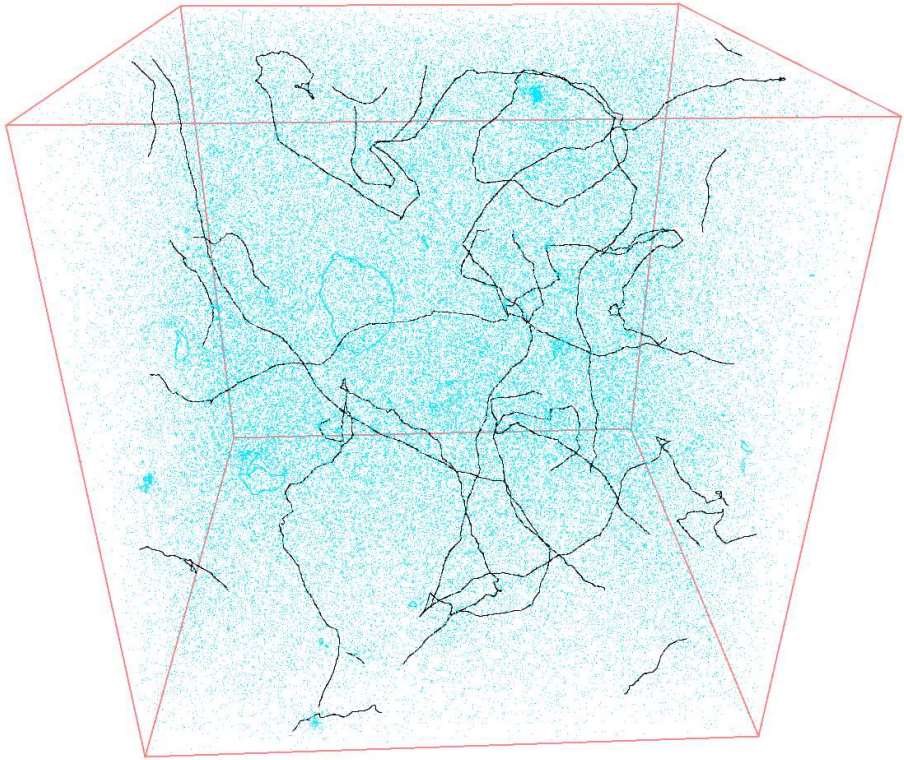}
    \caption{String network configuration in the matter era when the
      horizon fills the whole simulation box. String loops with a
      length smaller than the distance to the horizon appear in blue.}
    \label{fig:endmat}
  \end{center}
\end{figure}

By switching on the evolution from the initial network, string motion
and intersections drastically change the shape of the strings as well
as the network aspect (see Fig.~\ref{fig:endmat}).
Naively, without any collisional process, one would expect the string
network to dominate the energy density of the universe. In a volume
$V$, denoting by $\ellinf$ the typical correlation length of the network
at a given time (initially $\ellinf=\corrini$), the number of strings
should be roughly given by $V/\ellinf^3$. The resulting energy density
should therefore be
\begin{equation}
\label{eq:rhowrong}
  \rhoinf \simeq \dfrac{V}{\ellinf^3}\times (U \ellinf) \times \dfrac{1}{V}= \dfrac{U}{\ellinf^2}\,.
\end{equation}
Due to cosmological expansion one has $\ellinf \propto a$ and $\rhoinf
\propto 1/a^2$. As noted by Kibble, this domination does not occur due
to intercommutation processes which allow the formation of loops. In
the so-called ``one scale model'', Kibble~\cite{Kibble:1976} assumes
that loops of typical size $\ellinf$ are formed at a rate equals to
$\ellinf^{-4}$ (for relativistic speeds, one expects one
intercommutation per string during the time $\ellinf$). As a result,
during a time interval $\delta t$, the energy density transferred to
loops is
\begin{equation}
  \label{eq:tokloops}
\delta \rho_{\infty \rightarrow \circ} \simeq \ellinf^{-4} \delta t \,
U \ellinf.
\end{equation}
From Eq.~(\ref{eq:rhowrong}), the energy density of strings which are
not loops verifies
\begin{equation}
\label{eq:evolrhoinf}
\frac{\ud \rhoinf}{\ud t} \simeq -2 H \rhoinf -
\frac{\rhoinf}{\ellinf}\,,
\end{equation}
where $\ellinf$ is a function of the cosmic time. Defining $C(t) =
\ellinf(t)/t$, this equation can be recast into
\begin{equation}
\label{autosiminfty}
\frac{1}{C} \frac{\ud C}{\ud t} \simeq -\frac{1}{2t}
\left(\frac{2+6w}{3+3w} - \frac{1}{C} \right),
\end{equation}
where the background cosmological fluid sourcing the universe expansion
has an equation of state $P=w \rho$. The constant solution
$C(t)
=(3+3w)/(2+6w)$ is an attractor for which
\begin{equation}
\label{eq:rhoscaling}
\rhoinf \propto \frac{U}{t^2} \propto \frac{U}{a^{3(1+w)}}\,.
\end{equation}
The energy density associated with strings which are not loops
``scales'' as matter in the matter era and radiation in the radiation
era. Therefore, it is prevented to dominate over the usual
cosmological fluids and cannot overclose the universe. Notice however
that the total energy density could still dominate the cosmological
dynamics if the energy density under the form of loops is not
evacuated by some extra-mechanism. For NG cosmic strings, loops are
transformed into radiation due to the emission of gravitational
waves~\cite{Vilenkin:1981, Damour:2001bk, Siemens:2006vk,
  Olmez:2010bi}.  Other types of loops may loose energy by different
radiative processes, such as particle emission, or even energy leakage
into the extra-dimensions in the case of cosmic
superstrings~\cite{Vachaspati:2008su, Vachaspati:2009kq}.
\begin{figure}
  \begin{center}
    \includegraphics[width=12cm]{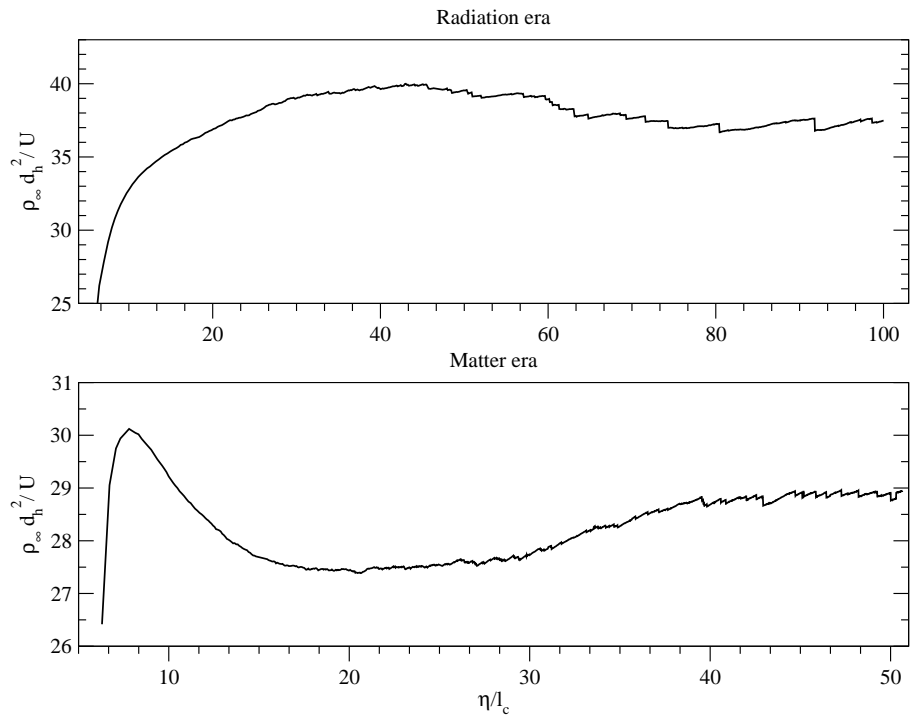}
    \caption{Energy density of super-horizon sized strings as a
      function of the conformal time (in unit of $\corrini$) in the
      radiation and matter era. After some relaxation time, $\rhoinf
      \propto 1/\horizon^2$, the proportionality factor is universal.}
    \label{fig:scalinf}
  \end{center}
\end{figure}
In Fig.~\ref{fig:endmat}, we have represented an evolved string
network at the end of a matter era run. For FLRW simulations within a
fixed comoving box with periodic boundary conditions, one cannot
evolve the system indefinitely: at some point, periodic boundaries
become causally connected. Usually, one stops the run when the
distance to the horizon fills the simulation volume, or more
rigorously half of it. In this figure, one sees that only a few
super-horizon strings remain (black long strings) whereas the box is
also filled with a lot of small loops (blue) and a few larger loops
having a size typical of distance between two long strings. The latter
are freshly formed Kibble loops whereas the existence of the small
ones cannot be explained in the framework of the one scale
model~\cite{Bennett:1990}. Concerning the energy density associated
with the super-horizon strings (also called infinite strings), their
evolution in the matter and radiation era have been plotted in
Fig.~\ref{fig:scalinf}: they ``scale'' as expected. From
Ref.~\cite{Ringeval:2005kr}, one has
\begin{equation}
\label{eq:infscaling}
\left. \rhoinf \dfrac{\horizon^2}{U}\right|_{\umat} = 28.4 \pm 0.9,
\qquad \left.\rhoinf \dfrac{\horizon^2}{U}\right|_\urad=
37.8 \pm 1.7.
\end{equation}
As the behaviour of the energy density associated with long strings
suggests, the time evolution drives the string network towards a
stable cosmological configuration which does not seem to depend on its
initial configuration, at least for the long
strings. Fig.~\ref{fig:endmat} therefore displays what a cosmological
string network should look like inside a horizon volume, at any time
during the matter era. As is clear from Fig.~\ref{fig:scalinf}, the
relaxation time required for the energy density of long strings to
reach the attractor is small. Concerning the cosmic string loops,
their existence and behaviour have been the subject of various claims
and analytical works~\cite{Austin:1993rg, Copeland:1998na,
  Ringeval:2005kr, Olum:2006ix, Martins:2006, Polchinski:2006ee,
  Dubath:2007mf, Copeland:2009dk}. In the following, we present recent
results~\cite{Ringeval:2005kr, Rocha:2007ni} showing that the energy
density of loops also reaches a scaling evolution similar to
Eq.~(\ref{eq:infscaling}).

\subsubsection{Loops.}
\label{sec:loops}
\begin{figure}
  \begin{center}
    \includegraphics[width=12cm]{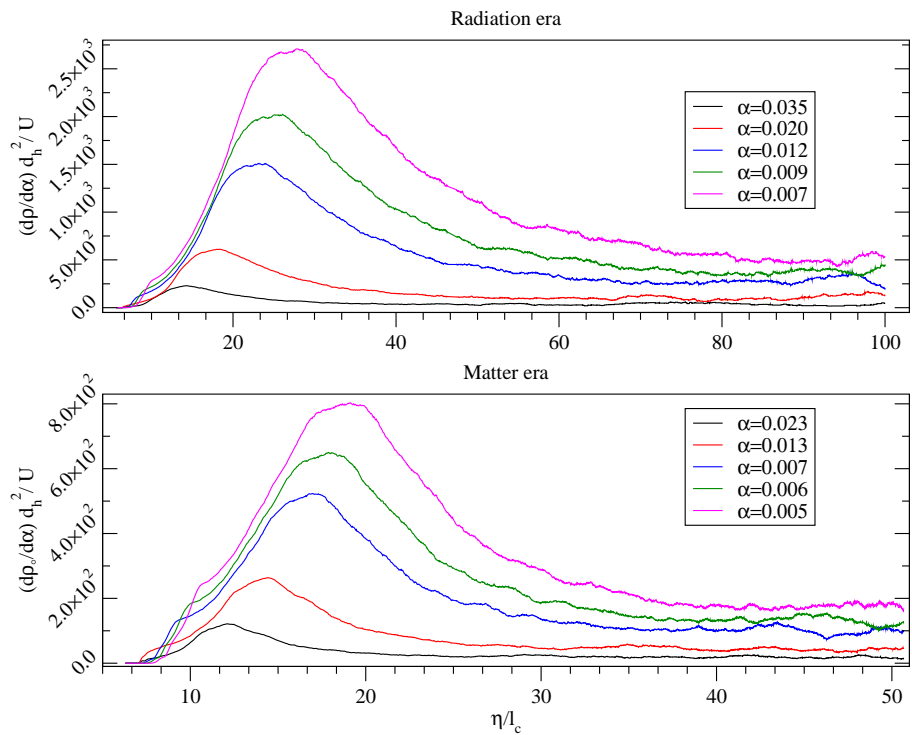}
    \caption{Evolution of the energy density distribution of $\frachor
      \horizon$-sized loops as a function of the conformal time in the
      radiation and matter era. After an overproduction regime, the
      energy density distribution scales as $1/\horizon^2$, as the
      long strings.}
    \label{fig:scalloop}
  \end{center}
\end{figure}
As previously mentioned, the small loops observed in NG simulation
cannot be explained in the framework of the one scale model. These
loops find their origin from the self-intercommutation of strings on
length scales typical of their small size. The building of a small
scale structure on strings is the outcome of the successive
intersection events during which new kinks are produced and propagate
along the intercommuted segments. Correlations between the kinks
induce, from Eqs.~(\ref{eq:bbkinks}), auto- and cross-correlations
between $\vecXd(\sigma_1)$ and $\vecXp(\sigma_2)$ from which small
loops can be produced~\cite{Austin:1993rg, Polchinski:2006ee}. In
Fig.~\ref{fig:scalloop}, we have plotted the energy density
distribution under the form of loops with respect to the conformal
time during the radiation and matter era. The simulation performed is
one of the largest up to date: the box contains $100\corrini$ whereas
the redshift simulation range reaches almost two orders of
magnitude. The loop energy density distribution $\ud \rholoop/\ud
\frachor$ is defined such that $\ud \rholoop(\frachor)$ is the energy
density carried by all loops having a physical length $\ell$ in the
range $\frachor \horizon$ to $(\frachor + \ud \frachor)\horizon$. In
other words, we measure loop size in unit of the horizon
length\footnote{This is the relevant physical length scale of the
  problem.}. A logarithmic binning in $\frachor$ of resolution $\Delta
\frachor/\frachor \simeq 10^{-1}$ has been used in the range
$\left[10^{-5},10^{2}\right]$ to compute these quantities. From this
plot, it is clear that after an overproduction regime characterized by
the bump of Fig.~\ref{fig:scalloop}, the energy density distribution
of loops of given size $\alpha$ relaxes towards a stationary regime in
which it scales as $1/\horizon^2$. Such an observation implies that,
once relaxed, the loop number density distribution is of the form
\begin{equation}
\label{eq:scaling}
\dfrac{\ud\numloop}{\ud
  \frachor} = \dfrac{\scaling(\frachor)}{\frachor \, \horizon^3}\,,
\end{equation}
where the ``scaling function'' $\scaling(\frachor)$ is found to be well
fitted by the power laws~\cite{Ringeval:2005kr}
$\scaling(\frachor) = \const \,
\frachor^{-\power}$ with
\begin{equation}
\label{eq:powerlaw}
\begin{array}{cc}
\left\{
\begin{array}{ccc}
\power &= & 1.41\,^{+0.08}_{-0.07} \\
\const & = & 0.09\,^{-0.03}_{+0.03}
\end{array}
\right|_\umat,
\quad \mathrm{and} \quad
\left\{
\begin{array}{ccc}
\power & = & 1.60\,^{+0.21}_{-0.15} \\
\const & = & 0.21\,^{-0.12}_{+0.13}
\end{array}
\right|_\urad,
\end{array}
\end{equation}
for the matter and radiation era, respectively.
\begin{figure}
  \begin{center}
    \includegraphics[width=10cm]{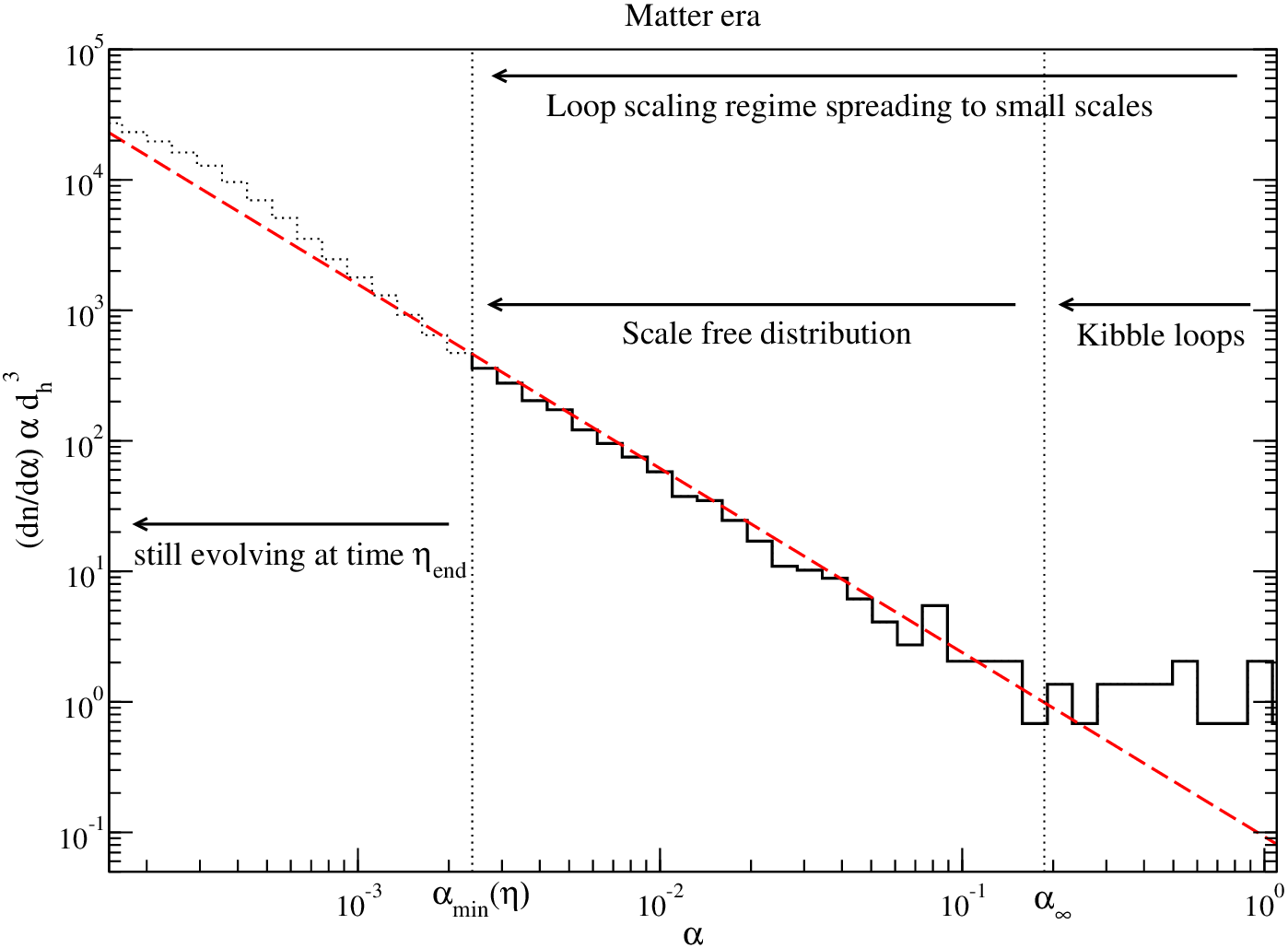}
    \caption{Loop number density distribution at the end of a
      $(100\corrini)^3$ matter era numerical simulation. Apart for a
      few Kibble loops ($\frachorlong<\frachor<1$), all loops of
      size $\ell=\frachor \horizon$ with $\frachor>\frachorscal$
      follow a scaling regime in which their number density is a power
      law. Smaller loops are still in the relaxation regime and will
      enter their scaling regime later (see
      Fig.~\ref{fig:relaxloop}). The red dashed line is the best power
      law fit of the scaling function $\scaling(\frachor)$ given by
      Eq.~(\ref{eq:powerlaw}).}
    \label{fig:numdens}
  \end{center}
\end{figure}
The loop number density distribution, for the matter era run, has been
plotted in Fig.~\ref{fig:numdens}. As Fig.~\ref{fig:scalloop} already
shows, the loop distribution takes more time to reach the scaling
regime for the small loops. The relaxation bump is all the more so
high and long than $\frachor$ is small. In the loop number density
distribution, this effect appears as a minimal time decreasing value
$\frachorscal(\eta)$ such that the loop distribution is in scaling at
$\frachor>\frachorscal$. The redshift range probes by a FLRW string
simulation is typically $\Delta z \simeq 10^2$, while for strings
formed at the GUT energy scales, one expects a $\Delta z\simeq
10^{18}$ at nucleosynthesis. It is clear that, in the cosmological
context, the string network has quite a time to relax: on all of the
relevant observable length scales the loop distribution should be in
scaling, i.e. $\frachorscal \ll 1$.  Since a power law distribution is
scale free, one concludes that a cosmologically stable string network
\emph{does not} exhibit loops of a particular size: this is not
surprising since the only length scale involved is the distance to the
horizon. These numerical results can be analytically recovered in the
framework of the Polchinski--Rocha model~\cite{Polchinski:2006ee}. The
expected loop number density distribution have been explicitly derived
by Rocha in Ref.~\cite{Rocha:2007ni} with a predicted power $p=1.5$
for the matter era and $p=1.8$ in the radiation era. If not due to
statistical errors, these small differences may be explained by the
existence of additional fractal micro-structure along the strings not
considered in the analytical approach (see also
Sec.~\ref{sec:smapng}). Of course, the previous statements hold
provided the other physical effects which are not included in the
simulation do not enter the game, as gravitational radiation and
gravitational back-reaction. The typical length scales at which they
should play a role is typically a multiple of $GU$, or some positive
power of it~\cite{PhysRevD.45.1898, Siemens:2002dj}. As shown in
Ref.~\cite{Rocha:2007ni}, gravitational radiation indeed cures the
energy density divergence that one can extrapolate from
Fig.~\ref{fig:numdens} when $\frachor \rightarrow 0$. Let us finally
notice that although the long strings are defined by $\frachor>1$,
there is also a small population of Kibble loops. Their typical size
being the horizon-sized correlation length of the long string network,
they can be defined to be those having $\frachorlong<\frachor<1$,
where
\begin{equation}
  \frachorlong=\dfrac{1}{\horizon} \left( \frac{U} {\rhoinf}
  \right)^{1/2}.
\end{equation}

\subsection{Relaxation towards scaling: memory of the initial
  conditions}

\begin{figure}
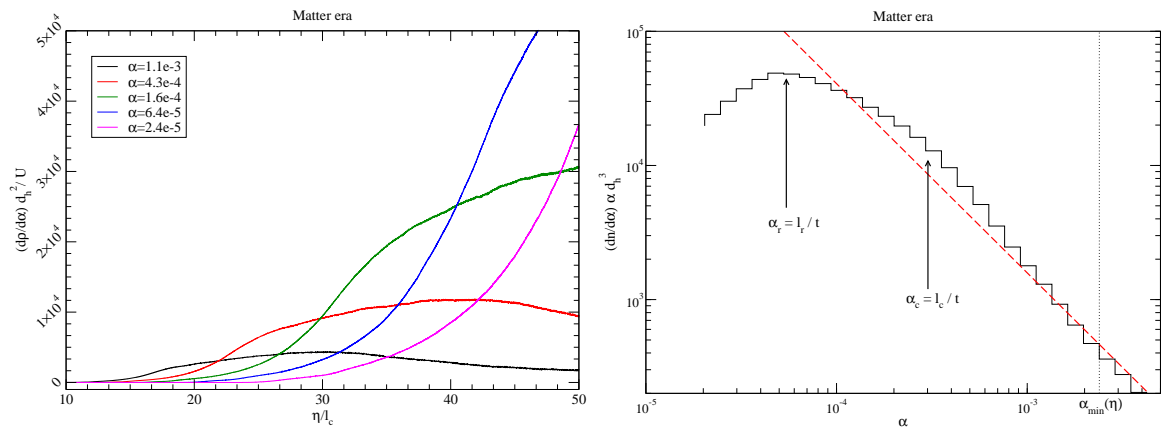

  \begin{center}
    \includegraphics[width=7.7cm]{nonscalmat}
    \includegraphics[width=7.5cm]{numrelaxmat}
    \caption{Loop energy density distribution with respect of the
      conformal time for the smallest loops. By the end of the
      numerical simulation, these loops have not yet reached their
      scaling behaviour (left panel). The right panel is the
      corresponding loop number density distribution at the end of the
      simulation ($\eta/\corrini=50$). Compared to the asymptotic
      scaling power law (red dashed), the loops in the relaxation
      regime are preferentially produced at fixed physical lengths
      given by the initial correlations present in the VV network.}
    \label{fig:relaxloop}
  \end{center}
\end{figure}

Although of less cosmological interest, the relaxation of the loop
energy density distribution towards its scaling regime shows
interesting properties which could explain some of the differences
observed between Abelian Higgs field simulation and NG simulations. In
the left panel of Fig.~\ref{fig:relaxloop}, we have plotted the loop
energy density distribution for loops smaller than the ones in
scaling. At the end of the numerical simulation, these length scales
are those having $\frachor<\frachorscal$.
This plot shows that the formation of the smallest loops is a delayed
mechanism which suggests that a cascading process takes place from the
initial string network configuration. For $\frachor\simeq 10^{-5}$,
only the increasing part of the relaxation bump appears at the end of
the simulation whereas the decaying towards scaling is still visible
for the larger $\frachor$. On the right panel, we have plotted the
loop number density distribution at the end of the simulation, i.e. at
the time $\eta/\corrini=50$ in the left panel. The loop number density
distribution deviates from the asymptotic scaling distribution on two
typical length scales.

Firstly, an overdensity compared to scaling is situated at
$\frachorini = \corrini/\horizon \propto 1/t$. In other words, there
is an overproduction of loops with a typical size equals to the
initial correlation length of the string network. Although one expects
the system to retain some memory of the initial conditions during the
relaxation, it may appear surprising that, in spite of the expansion
of the universe, the physical length scale of these loops remains the
same. A physical interpretation is that $\corrini \ll \horizon$ which
suggests that, at those small length scales, the system decouples from
the Hubble flow. More quantitatively, this effect can be explained in
the context of the three scale models~\cite{Austin:1993rg}. Under some
assumptions, the string small scale correlations can indeed sustain a
constant physical length.

Then, there is the overall peak of the loop number density
distribution at $\frachorres=\resini/\horizon \propto 1/t$. Most of
the loops which are not in scaling have this size at the end of the
simulation. In fact, one can check that these loops start appearing
soon after the beginning of the string evolution. This length scale
is, again, at a constant physical length $\resini$ and is associated
with a purely numerical effect~\cite{Ringeval:2005kr}. As discussed, a
numerical string is discretised with $\Nppcl$ points. The
Bennett--Bouchet code at the basis of the simulations presented here
uses an adaptive griding algorithm meaning that loops of any physical
size can be formed~\cite{Bennett:1990}. The only restriction is that,
at a given time, a loop is an object of at least three
points. Consequently, when the initial string network starts its
evolution, loops smaller than $\resini=\corrini\times 3/\Nppcl$ cannot
be formed. The existence of a finite numerical resolution therefore
adds some unwanted initial correlations of length $\resini$. Notice
that this is not a cut-off but indeed an extra-correlation.

As a basic consequence, one should not trust a NG simulation at those
length scales. However, the fact that the initial string network
violently relaxes towards scaling by emitting loops at the smallest
available correlation lengths has still some physical
significance~\cite{Vincent:1996rb}. What happens if we increase
$\Nppcl\rightarrow \infty$? As discussed in
Ref.~\cite{Ringeval:2005kr}, the larger length scales are not affected
and only the overall peak is shifted around the new $\resini
\rightarrow 0$. At these length scales, it is clear that using a NG
string to describe a network of topological defects would break down
and a reasonable assumption is to assume that the network will now
relax by losing energy through the relevant physical mechanism
available at those $\resini \rightarrow 0$ distances. In Abelian Higgs
simulation, most of the network energy is emitted through field
radiation, up to the point that almost no loop are observed in the
simulations~\cite{Vincent:1998,Hindmarsh:2008dw}. Abelian simulation
suffers from low resolution compared to NG ones and this has been a
subject of debate to decide whether or not this could explain the
absence of loops~\cite{Moore:2002, Hindmarsh:2008dw}. The above NG
results clearly support that particle and/or gravitational waves
emission is an important mechanism which certainly dominate the
relaxation regime. However, when the scaling regime progressively
takes place, from large to small length scales, the loop formation
mechanism should become dominant. One may speculate that it is not
clearly observed in Abelian simulations due to its delayed
appearance\footnote{A hint for this is the compatibility of the
  Abelian Higgs string correlators with the Polchinski--Rocha
  model~\cite{Hindmarsh:2008dw}; this one explaining the NG loop
  distribution~\cite{Rocha:2007ni}.}, as it is actually the case in NG
simulations when we compare it to the formation of $\resini$-sized
loops.

\section{Cosmic microwave background anisotropies and
  non-Gaussianities}
\label{sec:smapng}

At this point, numerical simulations of cosmic string evolution give
us the means to derive observable predictions. As should be clear from
the previous section, some structures in the numerical simulations are
not supposed to be present after a realistic expansion factor of
$10^{18}$: these are the structures still in the relaxation regime and
such that $\frachor < \frachorscal$, at any simulation time. In the
following, we will denote by ``systematic errors'', the uncertainties
inherent to the presence of such non-scaling structures when deriving
observable predictions from NG numerical simulations.

\subsection{Unequal time correlators}
The first method used to derive CMB anisotropies has been introduced
in Ref.~\cite{Zhou:1996} and applied in Refs.~\cite{Zhou:1996,
  Turok:1996ud, Pen:1997, Durrer:1997ep, Durrer:1998rw, Contaldi:1999,
  Wu:2002} for global topological defects and recently in
Ref.~\cite{Bevis:2007, Bevis:2010gj} for the Abelian strings.
\begin{figure}
  \begin{center}
    \includegraphics[width=9cm]{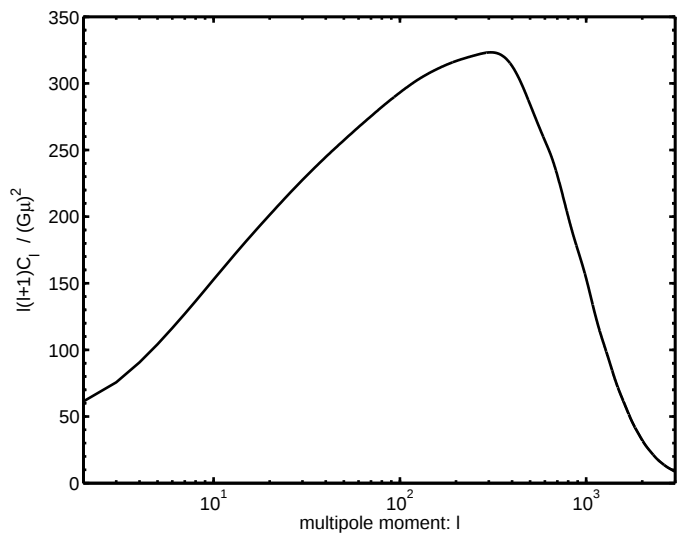}
    \caption{CMB temperature power spectrum induced by a network of
      Abelian cosmic strings and derived from the unequal time
      correlators method. Figure extracted from
      Ref.~\cite{Bevis:2006mj}.}
    \label{fig:TTAbelian}
  \end{center}
\end{figure}
Cosmic strings are active sources of gravitational
perturbations~\cite{Magueijo:1996px, Magueijo:1996xj} which means that
the equations of motion of their induced linear perturbation is of the
form
\begin{equation}
\calD \calP = \calS,
\end{equation}
where $\calD$ is a time differential operator, $\calP$ the
perturbation in the energy density, or velocity, etc\dots which is
directly related to the CMB temperature anisotropies. Here $\calS$
denotes the source terms, i.e. the string stress tensor. From the
Green's function $\calG$ of this equation, one gets, today (at
$\eta_0$) and in Fourier space
\begin{equation}
  \calP(k,\eta_0) \propto \int^{\eta_0} \calG_k(\eta) \calS(\eta,k) \ud \eta.
\end{equation}
The two-point correlator reads
\begin{equation}
  \langle \calP^\dag(\eta_0,k) \calP(\eta_0,k) \rangle \propto
  \int\int^{\eta_0}  \calG^\dag_k(\eta') \calG_k(\eta) \langle
  \calS^\dag(\eta',k) \calS(\eta,k) \rangle \ud \eta \ud \eta',
\end{equation}
and its determination requires a full-time knowledge of the source
term $\calS(\eta,k)$ for each mode. Since it is impossible to carry
out a simulation over the whole cosmological history, the scaling
properties of the cosmic string network can be used to analytically
extrapolate the source terms over the required ranges. As shown in
Refs.~\cite{Durrer:1997ep}, as long as the cosmic string network is in
a scaling regime, the source terms are the stress tensor components
and assume the form
\begin{equation}
\label{eq:uetc_scal}
\langle T_{\mu \nu}
(k,\eta) T_{\rho \sigma}^* (k,\eta') \rangle \propto \dfrac{1}{\sqrt{\eta
    \eta'}} f_{\mu \nu \rho \sigma} \left(k\sqrt{\eta \eta'},
  \dfrac{\eta}{\eta'} \right).
\end{equation}
Numerical simulations are actually used to determine the scaling
functions $f_{\mu \nu \rho \sigma}$. In Fig.~\ref{fig:TTAbelian}, we
have represented the temperature anisotropies derived in
Ref.~\cite{Bevis:2006mj} using such a method from Abelian Higgs string
simulations. The current CMB constraint on $GU$ comes from this power
spectrum: at two-sigma, $GU<7\times 10^{-7}$~\cite{Bevis:2007gh}.

\subsection{Simulated small angle maps}
\label{sec:simumaps}
The previous constraint typically corresponds to a string contribution
which cannot exceed $10\%$ at the multipole moment $\ell=10$. On
current observable angular scales, cosmic strings may only be a
sub-dominant fraction of the overall CMB anisotropies. However, string
induced perturbations being non-Gaussian, as opposed to inflationary
perturbations of quantum origin, one can go further than deriving the
two-point function. Notice that, in principle, the unequal time
correlator approach could be used to extrapolate the three- and higher
$n$-point function by using the scaling properties of the string
network. Another approach is to produce simulated maps of string
induced CMB anisotropies. Again, we face the problem of the small
redshift range probed by the numerical simulations. By putting an
observer inside the numerical simulation, such maps can only include
stringy effects up to a finite redshift, typically $z \simeq
10^2$. The CMB anisotropies computed in this way are therefore only
accurate on large angular scales but can produce full sky
maps~\cite{Pen:1994, Allen:1997, Landriau:2003, Landriau:2004}. This
limitation can be avoided by stacking maps from different redshifts,
an approach outlined in Ref.~\cite{Bouchet:1988} and applied in
Ref.~\cite{Bennett:1992,Fraisse:2007nu}.

Simulations with the observer outside of the numerical box are not
well suited for a full-sky map reconstruction, but are perfectly
designed for the small angular scales. The reason being that cosmic
strings are incessantly sourcing the CMB fluctuations since the last
scattering surface, and contrary to the perturbations of inflationary
origin, this part cannot be affected by Silk damping. Therefore, at
small angular scales, one expects the strings' signature in the CMB
temperature fluctuations to be dominated by their integrated
Sachs-Wolfe~(ISW) effect from the last scattering
surface~\cite{Fraisse:2007nu}.
\begin{figure}
  \begin{center}
    \includegraphics[width=10cm]{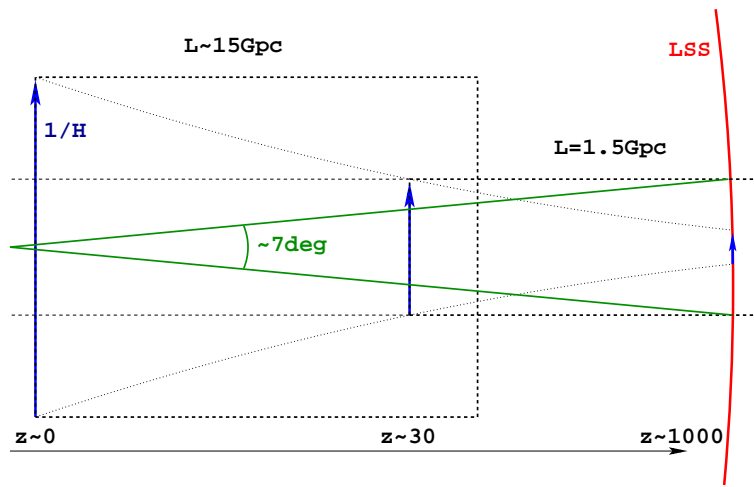}
    \caption{Stacking of two numerical simulations to compute the
      small angles CMB anisotropies induced by NG
      strings~\cite{Bennett:1992}. The dashed rectangles represent the
      redshift and angular extension of two numerical simulations used
      to evaluate Eq.~(\ref{eq:udef}). The first starts at last
      scattering and stops at $z\simeq 30$ while the second starts at
      $z\simeq 30$ and ends now, up to a small overlapping to ensure
      relaxation towards scaling.}
    \label{fig:stacking}
  \end{center}
\end{figure}
In the temporal gauge ($\tau=X^0=\eta$), the NG stress tensor derived
from Eq.~(\ref{eq:ngaction}) reads
\begin{equation}
\label{eq:ngstress}
T^{\mu\nu} = \dfrac{U}{\sqrt{-g}} \int \ud \sigma 
\left(\varepsilon\, \dot{X}^\mu \dot{X}^\nu - 
  \dfrac{1}{\varepsilon}\, \Xp^{\mu} \Xp^{\nu}\right)
\delta^3 \left(\vec{x} - \vec{X}\right).
\end{equation}
In the flat sky approximation, well suited for angles typically
smaller than the Hubble angular size at the epoch of interest,
Hindmarsh has shown that the ISW temperature anisotropies induced by
NG strings can be simplified to~\cite{Hindmarsh:1993pu, Stebbins:1995}
\begin{equation}
\label{eq:stgsa}
\FFTheta{\bk} \simeq 
  \dfrac{8\pi\mathrm{i}\, G U}{\bk^2}
  \int_{\vec{X}\,\cap\,\vec{x}_\gamma}
  \left(\bu \cdot \bk\right)
  \ue^{-\mathrm{i}\, \bk \cdot \bX}\,
  \varepsilon\, \ud \sigma,
\end{equation}
where $\Theta(\bx) \equiv \delta T(\bx)/\Tcmb$. The wave vector $\bk$
denotes the \emph{transverse} component of the three-dimensional
vector $\vec{k}$ with respect to the line of sight $\unitn$, whereas,
in the temporal gauge, $\vecu$ encodes the string stress tensor
distortions of the photon temperature and reads
\begin{equation}
\label{eq:udef}
\vecu = \vecXd - 
  \dfrac{(\unitn \cdot \vecXp) \cdot \vecXp}{1 +
  \unitn \cdot \vecXd}\,.
\end{equation}
As can be seen in Eq.~(\ref{eq:stgsa}), only the strings that
intercept the photon path $\vec{x}_\gamma$ can imprint their signature
in the CMB temperature fluctuations. The previous expression is
nothing but the Gott--Kaiser--Stebbins effect in the temporal
gauge~\cite{Kaiser:1984iv, Hindmarsh:1993pu, Gott:1990}. As a result,
the knowledge of $\vect{u}$, and therefore of the string trajectories
$\vect{X}$, is only required on our past light cone.  In the context
of string numerical simulations, the trajectories of all strings are
computed during all of the numerical simulation time. Therefore, to
compute $\vect{u}$, one only needs to determine which parts of the
string network intercept our past light cone and at what time.

In order to be able to generate a significant amount of maps using
such a method, it is more convenient to stack two medium sized
$(50\corrini)^3$ simulations along the lines sketched in
Fig.~\ref{fig:stacking}. The first one starts at the last scattering
surface and ends at a redshift fixed by the maximum expansion factor
achievable in the numerical box.  For the simulations we performed,
initially $\horizon \simeq 0.185$ (in unit of the fixed comoving
simulation volume), which corresponds to $1.7\,\Gpc$ and a field of
view of $\angfov \simeq 7.2^\circ$ (for the current fiducial
cosmological parameters~\cite{Komatsu:2008hk}). Such a run ends after
a 30-fold increase in expansion factor, corresponding to a redshift
$z \simeq 36$.  We then propagate the photons perturbed by the first
run into a second numerical simulation of the same size but starting
at $z_\ui\simeq 36$.  For another 30-fold increase in expansion
factor, this run ends at $z\simeq 0.3$.  As can be seen in
Fig.~\ref{fig:stacking}, the second simulation represents a much
larger real volume than the first one and therefore subtends a greater
angle in the sky.  As a result, only the sub-part of the second run
that matches the angle subtended by the first simulation is actually
used. As we will see later on, the CMB temperature maps are weakly
sensitive to the string network at low redshifts, simply because there
are almost no strings intercepting our past light cone in a recent
past, which makes this technique perfectly acceptable.
\begin{figure}
  \begin{center}   
    \includegraphics[width=7.cm]{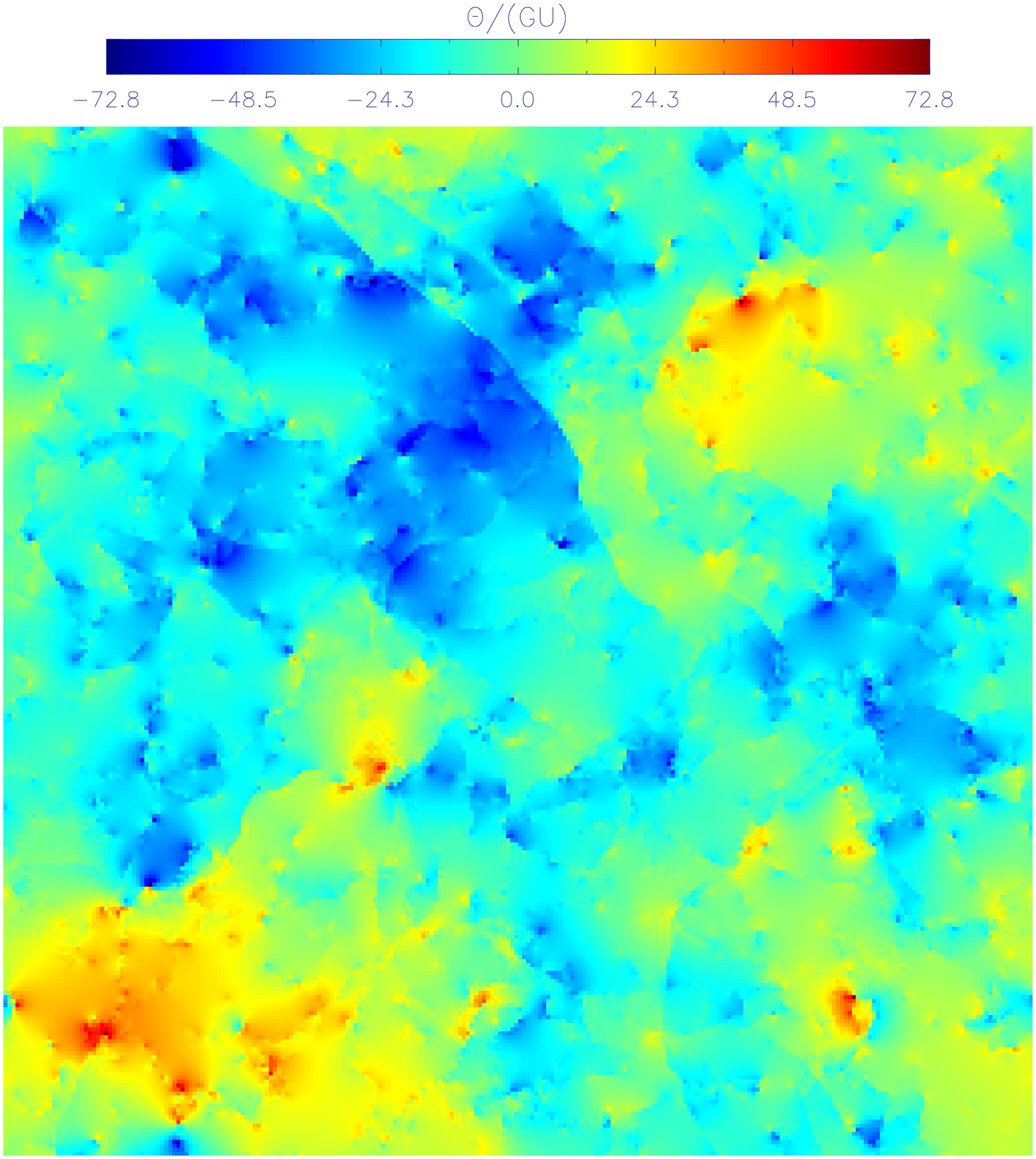}
    \includegraphics[width=7.cm]{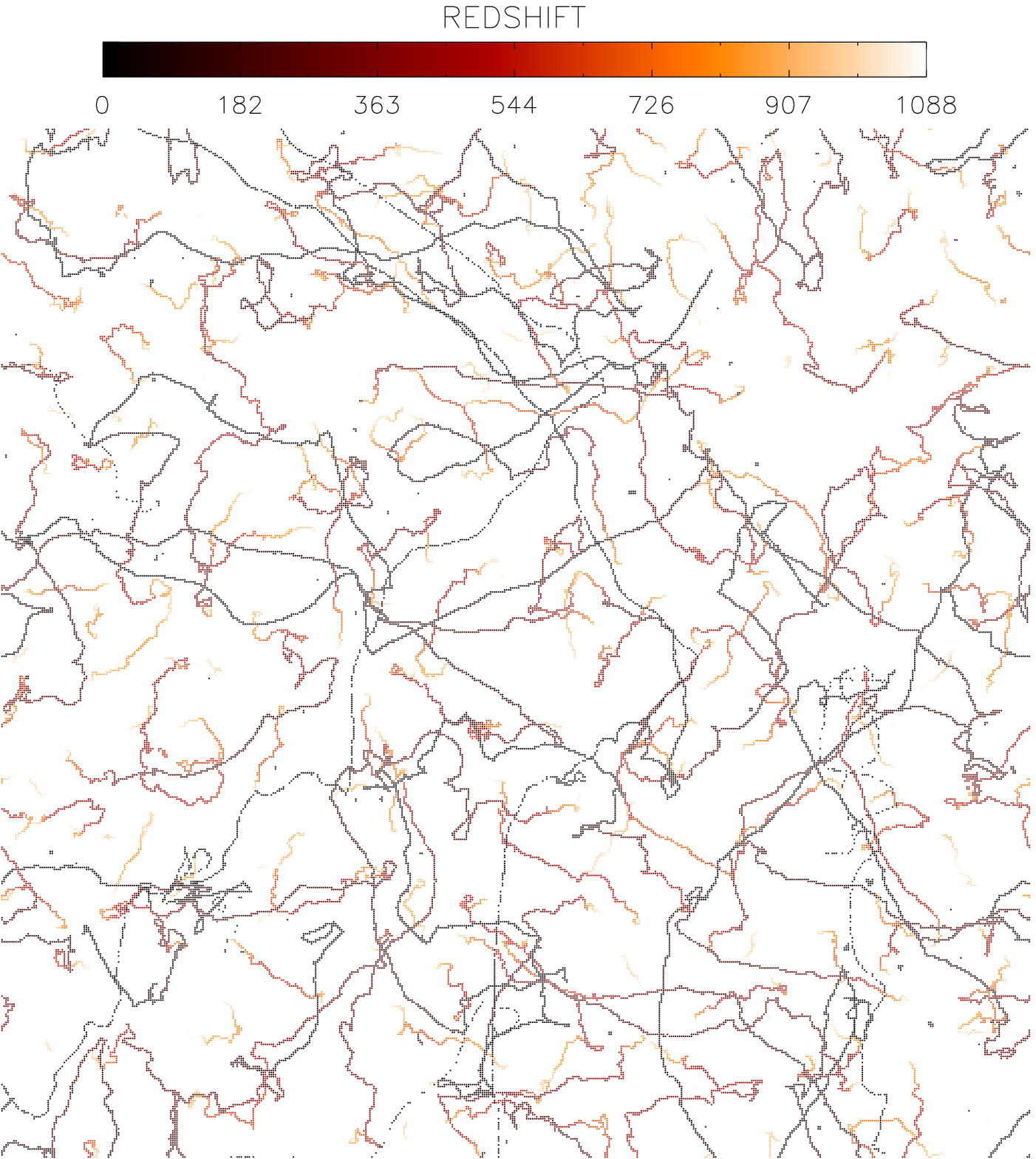}
    \caption{String-induced CMB temperature fluctuations on a
      $7.2^\circ$ field of view~\cite{Fraisse:2007}. Because of their
      cosmological scaling, most of the long strings intercept our
      past light cone close to the last scattering surface. As can be
      seen in the right image, the edges in the temperature patterns
      can be identified to strings intercepting our past light cone.
      Note that active regions corresponding to string intersection
      and loop formation events lead to the bright spots in these
      maps.}
    \label{fig:smap}
  \end{center}
\end{figure}
In practice, each of these numerical simulations is started before the
redshifts mentioned, in order to give the cosmic string network enough
time to relax toward its stable cosmological configuration. As
discussed in the previous section, one has to make sure that the
structures (strings and loops) we are interested in have indeed
reached their scaling behavior during the numerical runs. We switch on
the photon propagation inside the runs only after making sure all the
large structures (infinite strings and loops) are in their scaling
regime. This can be checked by monitoring the evolution of the energy
density distributions, and we have chosen to start the photons'
propagation when all loops larger than a third of the horizon size are
in scaling. This cutoff is then dynamically pushed to smaller values
to include all the loops entering the scaling regime at later
times. The cutoff time dependence is simply the function
$\frachorscal(\eta)$ and can be deduced from the loop distribution
relaxation times derived in Sec.~\ref{sec:loops}. The resulting CMB
temperature map is displayed in the left panel of Fig.~\ref{fig:smap}
whereas the right panel shows the string paths projected onto our past
light cone. Again, these maps are only representative at small angular
scales. On larger angles, they represent only the ISW contribution to
the overall string anisotropies: for instance, Doppler contributions
coming from photon decoupling at last scattering are dominant around
$\ell=300-400$ (see Fig.~\ref{fig:TTAbelian}).

The discussion on systematic effects coming from the presence of loops
not yet in scaling can be found in Ref.~\cite{Fraisse:2007}. In fact,
they have only a small effect. The physical reason being that, due to
scaling, the long strings are still the main source of CMB
anisotropies even at (reasonably) small angles. Indeed, there are
always roughly ten strings per Hubble volume at each time, which means
that a patch of $0.8^\circ$ is at least crossed by the ten long
strings being there at last scattering, plus a few others from lower
redshifts.

\subsection{Skewness and kurtosis}

\begin{figure}
\begin{center}
\includegraphics[width=9cm]{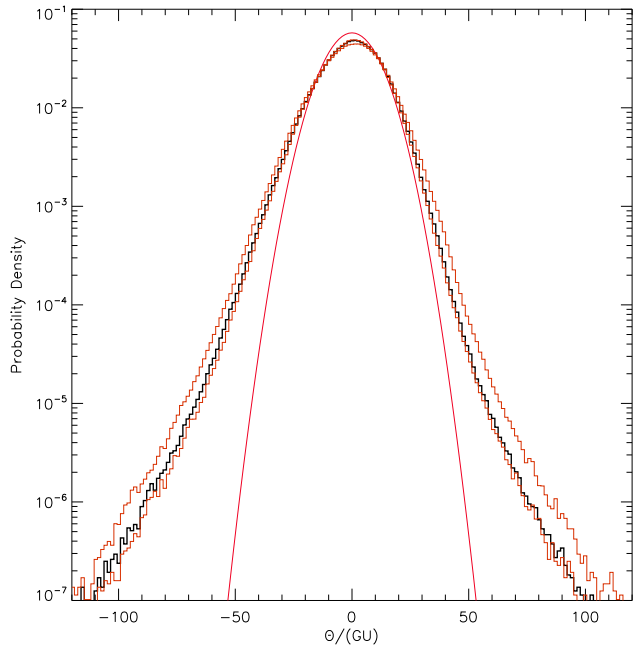}
\caption{The probability distribution function of CMB temperature
  fluctuations induced by NG cosmic strings. The orange curves
  quantify systematic errors coming from the string simulations by
  including non-scaling loops, or by removing all loops. Deviations
  from Gaussianity are clearly apparent in the tails of the
  distribution, as well as from the negative skewness.}
\label{fig:1pt}
\end{center}
\end{figure}
The most basic statistical test that can be performed from a set of
small angle CMB maps is to plot the one-point function of the
temperature anisotropies. As can be seen in Fig.~\ref{fig:1pt}, the
temperature anisotropies induced by cosmic strings are clearly
non-Gaussian. From a set of $300$ independent CMB maps, one finds the
mean sample skewness to be negative
\begin{equation}
\label{eq:skew}
g_1 \equiv \left \langle
\frac{\overline{(\Theta-\bar{\Theta})^3}}{\sigma^3} \right \rangle
\simeq -0.22 \pm 0.12,
\end{equation}
where the brackets stand for the mean over different realisations
while the bar denotes averaging on each map. The variance itself
averages to
\begin{equation}
\label{eq:var}
\sigma^2 \equiv \left\langle \overline{(\Theta -
 \bar{\Theta})^2}\right \rangle\simeq   \left(150.7 \pm 18 \right) (GU)^2.
\end{equation}
The quoted errors are statistical and refer to the square root of the
variance between the different realisations. Similarly, the mean
kurtosis averages to
\begin{equation}
g_2 \equiv
\left<\frac{\overline{(\Theta-\bar{\Theta})^4}}{\sigma^4}\right> -3 
\simeq 0.69 \pm 0.29. 
\end{equation}
An analytical approach extending these results to cosmic superstrings
can be found in Ref.~\cite{Takahashi:2008ui}. A simple way to look for
strings is to search for large (but rare) temperature
fluctuations. Deviations from Gaussianity start to be significant, let
us say by a factor of two, only in the tails when the probability
distribution becomes typically lower than $10^{-6}$.

\subsection{Real space methods}

Strings induce step-like discontinuities in the CMB anisotropies and
various methods have been designed to probe the non-Gaussianities
associated with them.

Multifractal analysis~\cite{Falconer:fracgeo} has the advantage of
being directly applicable to the time-ordered data retrieved when a
CMB telescope scans the sky. As opposed to the fractal dimension of a
set which measures how sparse it is, the multifractal spectrum of a
measure defined over a set gives how many and which fractal dimensions
there are. In the context of cosmic strings, this method has been
applied in Ref.~\cite{1995ApJ4491P} on one dimensional scans of maps
similar to the one in Fig.~\ref{fig:smap}, the measure being defined
by
\begin{equation}
\label{eq:scanmeas}
\mu(i) = \left[\Theta(i)-\Theta(i+1)\right]^2,
\end{equation}
where the integer $i$ labels a point along the scan. The multifractal
properties of this measure have been shown to be distinctive enough to
detect strings, compared to a Gaussian signal, but only when the
detector resolution is sufficiently good. One may wonder, under
multifractality, how a non-Gaussian string pattern could be
distinguished from other non-Gaussian sources. In fact,
Eq.~(\ref{eq:scanmeas}) consists in taking the gradient of the induced
CMB fluctuations along the scan. Step-like discontinuities, passed
over a gradient filter, become one-dimensional delta functions, and
this is a definite string feature that can only be altered by the beam
experiment.
\begin{figure}
\begin{center}
\includegraphics[width=7.cm]{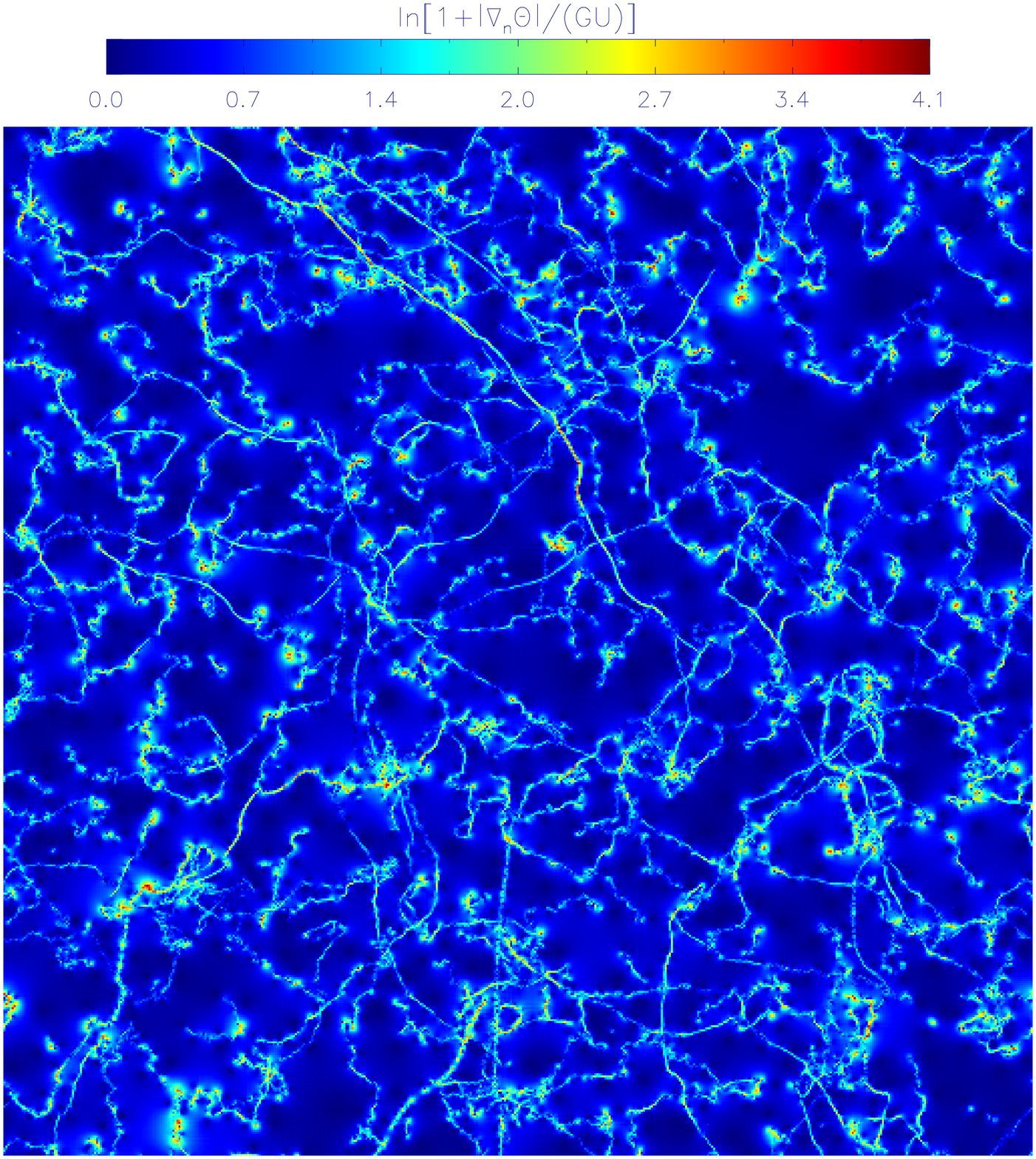}
\includegraphics[width=7.cm]{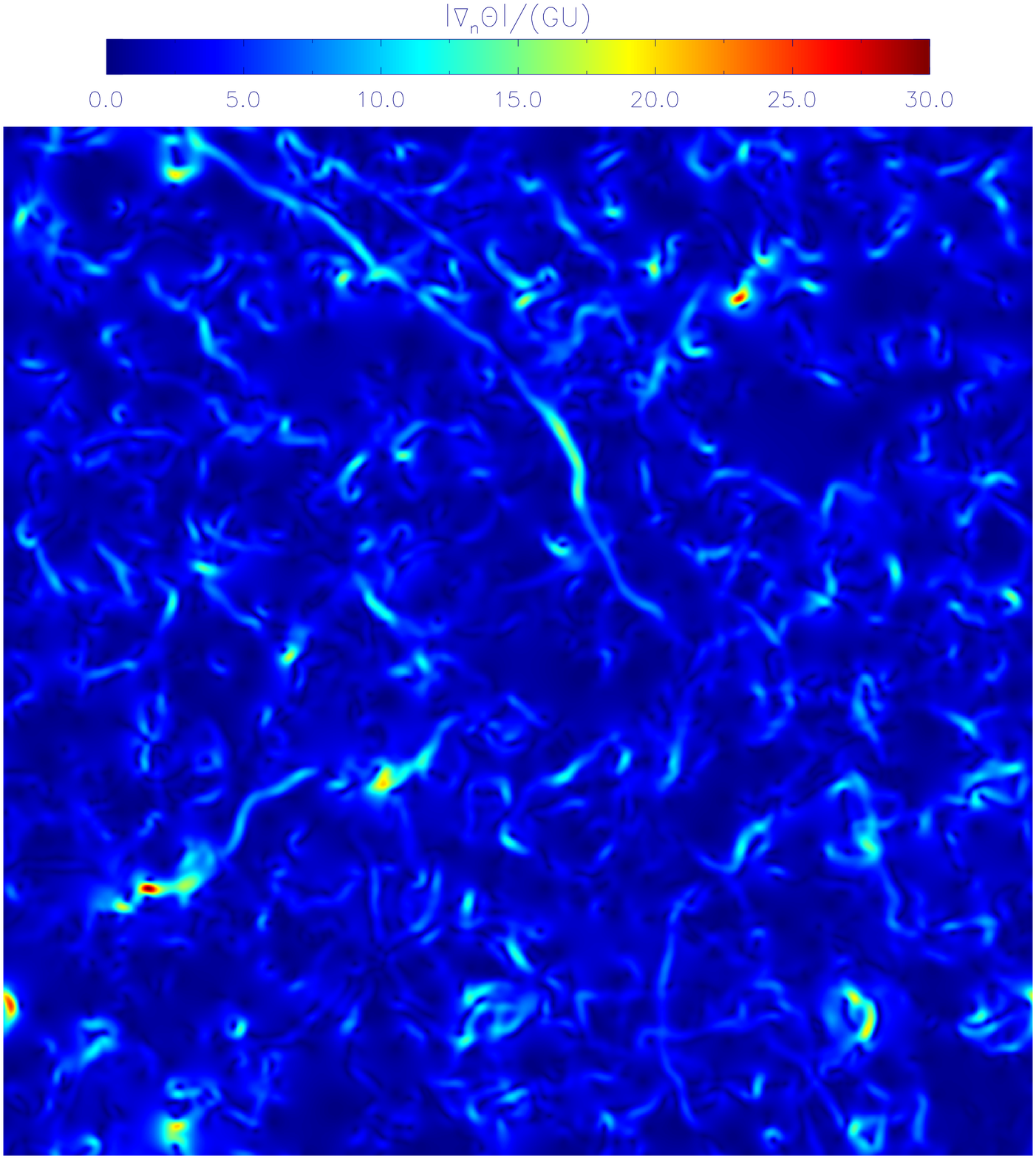}
\caption{Normalized gradient magnitude of the string-induced
  temperature anisotropies shown in Fig.~\ref{fig:smap} (left). A
  logarithmic scale has been used to enhance the contrast by
  preventing the bright spots from saturating the color scale. The
  right panel is the gradient magnitude obtained after convolution by
  a Planck-like Gaussian beam of resolution $5'$. Notice that the
  color scale is back to linear, most of the bright spots being now
  smoothed by the beam.}
\label{fig:grad}
\end{center}
\end{figure}
Denoting by $\angx$ and $\angy$ the horizontal and vertical angular
coordinates, the gradient magnitude $|\nabla \Theta|$ of the
temperature anisotropies is defined by
\begin{equation}
\label{eq:gradmag}
\left| \nabla \Theta \right| \equiv \sqrt{
  \left(\dfrac{\ud \Theta}{\ud \angx}\right)^2 +
  \left(\dfrac{\ud \Theta}{\ud \angy}\right)^2}\, .
\end{equation}
This definition makes it clear that for a finite temperature step,
let's say $\Theta(\angx,\angy) = \Theta_0\,\heaviside{\angx-\angx_0}$,
$\mathrm{H}$ being the Heaviside function, the resulting gradient
magnitude is a Dirac distribution at the string location. In
Fig.~\ref{fig:grad}, we have plotted the gradient magnitude of the
temperature maps of Fig.~\ref{fig:smap}, as well as its convolved
version with a Gaussian beam typical of the Planck satellite at
$217\,\GHz$. With a finite resolution beam, the discontinuities are
now smoothed. Real space methods applied to string are therefore
strongly sensitive to the angular resolution. Let us mention that
wavelet analysis methods have been also explored in this
context~\cite{1999MNRAS.309..125H, 2001MNRAS.327..813B} or to produce
cleaner maps~\cite{Hammond:2008fg}.

Directional gradients, obtained by variations with respect to either
$\angx$ or $\angy$, have been discussed in Ref.~\cite{Gott:1990} in
the context of Minkowski functionals. They are again found to provide
a more distinctive non-Gaussian behaviour than the original
temperature map for the contour length and genus.

\subsection{Temperature power spectrum}
\label{sec:cls}

\begin{figure}
  \begin{center}
    \includegraphics[width=9cm]{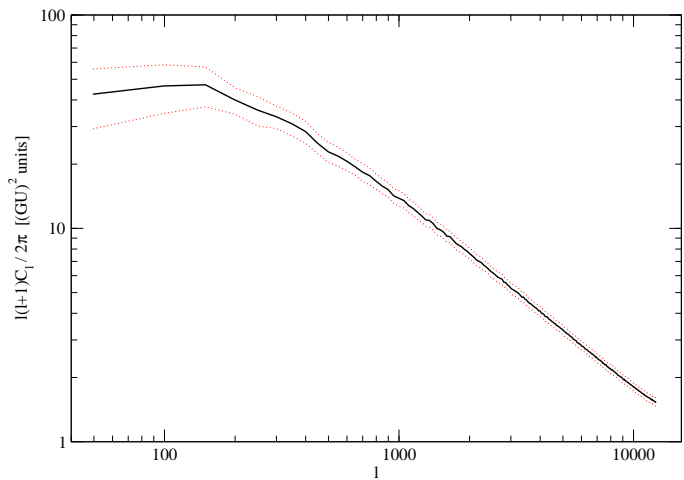}    
    \caption{Mean angular power spectrum of the string induced CMB
      anisotropies at small angular scales and its one sigma statistical
      errors (averaged over $300$ maps). At ``small'' multipoles
      $\ell<500$, Doppler contributions from the last scattering surface
      are expected to be significant and this plot gives only the ISW
      component~\cite{Fraisse:2007}.}
    \label{fig:clstg}
  \end{center}
\end{figure}

Moving to Fourier space, the small angle CMB maps also permit a
determination of the power spectrum at large multipoles. In
Fig.~\ref{fig:clstg}, we have plotted its mean value over the
different maps as well as the one-sigma statistical error around the
mean.
The overall power at $\ell=1000$ is~\cite{Fraisse:2007}
\begin{equation}
  \left.  \dfrac{\ell(\ell+1)\,C_\ell}{2\pi}\right|_{\ell=1000} 
  \simeq 14\,(GU)^2,
\end{equation}
which is close to the value obtained in Abelian Higgs field simulation
(see Fig.~\ref{fig:TTAbelian}). This is not so surprising since the
long strings in both NG and Abelian Higgs simulation have a similar
scaling evolution, and as explained above, long strings are the main
sources of CMB anisotropies even at the small angles. The power law
tail in Fig.~\ref{fig:clstg} is the direct consequence of the presence
of strings at all times since the last scattering surface: one finds
for $\ell \gg 1$~\cite{Fraisse:2007}
\begin{equation}
\label{eq:clsfit}
\ell(\ell+1)\, C_\ell  \propto \ell^{-p}
\quad \mathrm{with} \quad
p = 0.889^{+0.001}_{-0.090}\,,
\end{equation}
where only the systematic errors have been reported. Such a power law
shows that cosmic strings have to become the dominant primary source
of CMB anisotropies at the small angular scales, the fluctuations of
inflationary origin being killed by Silk damping at those multipoles.
\begin{figure}
  \begin{center}
    \includegraphics[width=9.3cm]{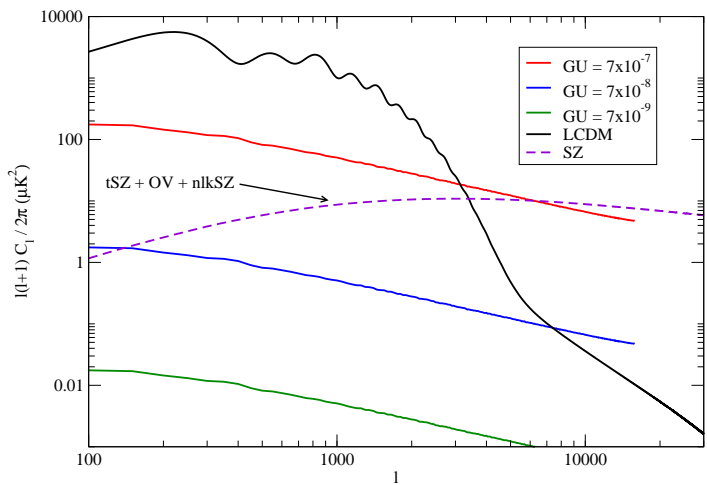}
    \caption{CMB temperature anisotropies from various sources,
      compared to the expected string contribution. Due to the Silk
      damping of primordial perturbations, string induced anisotropies
      always become dominant for the large multipoles. However, an
      unresolved SZ effect may compromise such a clean
      signature~\cite{Fraisse:2007}.}
    \label{fig:clsall}
  \end{center}
\end{figure}
In Fig.~\ref{fig:clsall}, we have plotted the respective contributions
of strings and adiabatic anisotropies of inflationary origin. The
cosmological parameters have been set to their fiducial values in the
Lambda-Cold-Dark-Matter (LCDM) model and the string energy density $U$
is compatible with the current bounds. For the current upper limit on
$GU=7\times 10^{-7}$, CMB anisotropies should become dominated by
cosmic strings at $\ell \gtrsim 3000$. An unresolved
Sunyaev--Zel'dovich (SZ) component may, however, compromise such a
signal. Nevertheless, string induced anisotropies do not depend on the
signal frequency whereas the SZ does, and one may hope to disentangle
both~\cite{Fraisse:2007}.

\subsection{Hindmarsh approximation}
\label{sec:hind}

As shown by Hindmarsh in Ref.~\cite{Hindmarsh:1993pu}, the power law
behaviour of the NG string power spectrum at small angles can be
analytically recovered. In Fourier space, the power spectrum of the
string induced CMB anisotropies $\Theta$ is defined by
\begin{equation}
\lvev{\FFTheta{\bk_1}\FFTheta{\bk_2}} = P(k_1)(2\pi)^2\delta(\bk_1+\bk_2),
\end{equation}
the expression of $\FFTheta{\bk}$ being given by
Eq.~(\ref{eq:stgsa}). Remembering that the string positions and
velocity vectors have to be evaluated on the past light cone, it is
more convenient to use the so-called light cone gauge. Instead of
identifying the timelike worldsheet coordinate $\tau$ with the
background time at the string event, one chooses instead to identify
$\tau=X^+ \equiv X^0 + X^3$. In this gauge, Eq.~(\ref{eq:stgsa})
simplifies to
\begin{equation}
  -k^2\FFTheta{\bk} = i \ep k_A   \int \ud\sigma \Xd^A(\sigma)
  e^{i\bk\cdot\bX(\sigma)},
\label{eq:lcgpert}
\end{equation}
where we have defined
\begin{equation}
  \ep = 8\pi GU,
\end{equation}
and where the capital indices are two-dimensional. The time parameter
$\tau$ then labels the intersections of a set of null hyperplanes with
the worldsheet. For our problem, all quantities have to be evaluated
at $\tau=x^+=\eta+z$. In a field of view of formal area
$\Area=(2\pi)^2\delta(0)$, one can express the power spectrum as
\begin{equation}
\label{eq:powermark}
  P(k)  = \ep^2 {k_Ak_B\over \Area k^4} \int \ud\sigma \ud\sigma' 
  \lvev{
    \Xd^A(\sigma)\Xd^B(\sigma') e^{i\bk\cdot
      [\bX(\sigma)-\bX(\sigma')]}}.
\end{equation}
Adding the assumptions that both $\Xd^A$ and $\Xp^B$ obey Gaussian
statistics, all of the correlation functions of $\FFTheta{\bk}$ can
now be written in terms of two-point functions only. Using the same
notation as in Ref.~\cite{Hindmarsh:1993pu}, the non-vanishing
two-point functions are
\begin{eqnarray}
  \label{eq:velocity}
  \lvev{\Xd^A(\sigma)\Xd^B (\sigma')} & = &\frac{1}{2} \delta^{AB} V(\sigma-\sigma'), \\
  \lvev{\Xd^A(\sigma) \Xp^B (\sigma')} & = &  \frac{1}{2} \delta^{AB} M (\sigma-\sigma'), \\
  \lvev{\Xp^A(\sigma) \Xp^B (\sigma')} &= & \frac{1}{2} \delta^{AB}
  T(\sigma-\sigma'),
\label{eq:tangent}
\end{eqnarray}
as well as the quantities
\begin{eqnarray}
  \Gamma(\sigma-\sigma') & \equiv& \lvev{\left[
      \vect{X}(\sigma)-\vect{X}(\sigma')\right]^2} = \int_{\sigma'}^\sigma
  \ud\sigma_1\int_{\sigma'}^\sigma \ud\sigma_2 T(\sigma_1-\sigma_2),\\
  \Pi(\sigma-\sigma') & \equiv &
  \lvev{\left[\vect{X}(\sigma)-\vect{X}(\sigma'))\right] \cdot \vect{\Xd}(\sigma')} = \int_{\sigma'}^\sigma \ud\sigma_1M(\sigma_1-\sigma').
  \label{eq:mixcor}
\end{eqnarray}
The leading terms are given by~\cite{Hindmarsh:1993pu,
  Hindmarsh:2009qk}
\begin{eqnarray}
\label{eq:LOcorr}
 V(\sigma) \to  \left\{
\begin{array}{cl} 
\vb^2 & \sigma \to 0 \\ 0 & \sigma \to \infty
\end{array}
\right.,\qquad
\Gamma(\sigma) \to  \left\{
\begin{array}{cl}
\tb^2\sigma^2 & \sigma \to 0 \\
   \corr\sigma & \sigma \to \infty
\end{array}
\right., \\
\Pi(\sigma) \to \left\{ 
\begin{array}{cl}
\half \frac{c_0}{\corr}\sigma^2 & \sigma \to 0 \\
0 & \sigma \to \infty
\end{array}
 \right. ,
\end{eqnarray}
where we have defined 
\begin{eqnarray}
\label{eq:projconsts}
  \corr = \Ga'(\infty), \qquad \vb^2 = \lvev{\bXd^2}, \qquad \tb^2 =
  \lvev{\bXp^2}, \qquad c_0 = \corr \lvev{\bXpp \cdot \bXd}.
\end{eqnarray}
The correlation length $\corr$ is the projected correlation length
on the past light cone, $\bar t^2$ is the mean square projected
tangent vector, $\bar v^2$ is the mean square projected velocity and
$c_0$ the correlation between projected velocity and curvature. From
these assumptions, Eq.~(\ref{eq:powermark}) reduces to
\begin{equation}
 P(k)  = \dfrac{\ep^2}{2\Area k^2}\int \ud\sigma \ud\sigma'\left[
   V(\sigma-\sigma') + \half k^2 \Pi^2(\sigma-\sigma')
 \right] e^{-k^2\Ga(\sigma-\sigma')/4}.
\label{ePowSpe}
\end{equation}
When $k\corr$ gets large, the terms involving the mixed correlator $M$
can be shown to be sub-dominant and only the first term remains:
\begin{equation}
  P(k) = \dfrac{\ep^2}{4 \Area k^2}  \int \ud\sigma_+ \ud\sigma_- V(\sigma_-)
  e^{-k^2\Ga(\sigma_-)/4},
\end{equation}
where $\sigma_\pm = \sigma\pm\sigma'$. Denoting by $L$ the total
transverse light-cone gauge length of string in the box of area
$\Area$, one gets
\begin{equation}
\label{eq:powersa}
 k^2P(k) \simeq  \ep^2  \sqrt{\pi} \frac{L\corr}{\Area} \frac{\vb^2}{\tb}
 {1\over (k\corr)}\, .
\end{equation}
At small angles, the wave number $k^2 \simeq \ell(\ell+1)$ and
Eq.~(\ref{eq:powersa}) predicts that $\ell(\ell+1)C_\ell \propto
\ell^{-1}$. The small difference with Eq.~(\ref{eq:clsfit}) is
suggestive of a cloud of zero-dimensional objects along the string
worldsheet which may be the signature of small loop production in the
NG numerical simulations. Let us stress that Eq.~(\ref{eq:powersa}) is
not ``primordial'' but directly approximate the observed angular power
spectrum of the CMB temperature anisotropies.

\subsection{Bispectrum}
\label{sec:bispec}

\subsubsection{Analytical approach.}

The success of Hindmarsh approximation to describe the small angular
CMB anisotropies power spectrum suggests it can be applied to higher
$n$-point functions. In Ref.~\cite{Hindmarsh:2009qk}, this method was
used to derive the bispectrum defined from the three points function
by
\begin{equation}
\vev{\FFTheta{\bk_1}\FFTheta{\bk_2} \FFTheta{\bk_3}} =
\BT(\bk_1,\bk_2,\bk_3)(2\pi)^2\delta(\bk_1+\bk_2+\bk_3).
\end{equation}
Plugging Eq.~(\ref{eq:lcgpert}) into the previous expression gives
\begin{equation}
  \BT(\bk_1,\bk_2,\bk_3) = i \ep^3 \frac{1}{\Area} \frac{k_{1_A}
    k_{2_B} k_{3_C}}{k_1^2 k_2^2 k_3^2}  \int
  \ud\sigma_1 \ud\sigma_2 \ud\sigma_3 \lvev{\Xd^A_1 \Xd^B_2 \Xd^C_3 
    e^{i \delta^{ab} \bk_a \cdot \bX_b}},
\end{equation}
with $\Xd^A_a = \Xd^A(\sigma_a)$, $a,b \in \{1,2,3\}$, and
$\bk_1+\bk_2+\bk_3=0$. With the Gaussian assumption, the ensemble
average of the string observables is lengthy but straightforward and
the final result reads~\cite{Hindmarsh:2009qk}
\begin{eqnarray}
&&\BT(\bk_1,\bk_2,\bk_3) \nonumber \\
&& = -\ep^3 \pi
  c_0\frac{\vb^2}{\tb^4}\frac{L\corr}{\Area}\frac{1}{\corr^2}
  \frac{1}{k_1^2k_2^2k_3^2} \left[ \frac{k_1^4
      \ka_{23} + k_2^4\ka_{31} + k_3^4\ka_{12}}{\left(\ka_{23}\ka_{31}
      +\ka_{12}\ka_{31}+ \ka_{12}\ka_{23} \right)^{3/2}} \right].
\label{eq:bistring}
\end{eqnarray}
The quantities $\ka_{ab}$ are shorthand for the scalar products
$\ka_{ab} \equiv -\bk_a \cdot \bk_b$. In the same way as for the power
spectrum, this expression directly gives the bispectrum of the CMB
temperature anisotropies. Its overall dependence varies as
$1/k^6$. Its sign depends on the sign of $c_0$ defined in
Eq.~(\ref{eq:projconsts}), and contrary to what one could naively
expect $c_0 \neq 0$: the projected string velocity and curvature
vectors are correlated. This can be shown by starting again from the
equations of motion (\ref{eq:ngtrans}), but this time, in the light
cone gauge. The equation of motion for $X^+$ gives
\begin{equation}
\frac{\dot\varepsilon}{\varepsilon} + 2 \calH(\Xd^0 + \Xd^2) = 0,
\end{equation}
whereas the equation for the transverse components is
\begin{equation}
  \bXdd + 2 \calH\frac{1}{\varepsilon^2}\left( \bXp^2\right) \bXd  -
  \frac{1}{\varepsilon} \frac{\partial}{\partial\sigma}\left(\frac{1}{\varepsilon}
    \frac{\partial
      \bX}{\partial\sigma} \right) - 2 \calH \frac{1}{\varepsilon^2}(\bXd\cdot\bXp) \bXp = 0.
\end{equation}
In a FLRW background, assuming that $\vev{\bXd^2}$ is constant, and
neglecting higher-order correlations between $\calH$, $\bXd$ and $\bXp$, we find
\begin{equation}
  \lvev{\frac{\partial^2\bX}{\partial s^2}\cdot\bXd}  = 2 \bar\calH \lvev{
    \left(\frac{\partial \bX}{\partial s}\right)^2\bXd^2}  - 2 \bar{\calH}
  \lvev{\left(\bXd\cdot\frac{\partial\bX}{\partial s}\right)^2} ,
\end{equation}
where we have defined $\ud s = \varepsilon \ud\sigma$, and where $\bar
\calH$ is the averaged conformal Hubble parameter. Still assuming that
the ensemble is approximately Gaussian in $\bXd$ and
$\bXp/\varepsilon$, the right hand side reduces to
\begin{equation}
\lvev{\dfrac{\partial^2 \bX}{\partial s^2} \cdot \bXd} = \bar{\calH}
\left( \lvev{\bXd^2} \lvev{\bXp^2} - \lvev{\bXd \cdot \bXp}^2 \right).
\end{equation}
The last term vanishes and the cross correlator $\bar{\calH} \vb^2
\tb^2$ is positive: from Eq.~(\ref{eq:projconsts}), we deduce that
$c_0 > 0$. It is interesting to notice that $c_0$ would vanish in
Minkowski spacetime, which can be viewed as a consequence of time
reversal invariance. The existence of a cosmic string bispectrum is
the consequence of the breaking of the time reversal invariance in a
FLRW background.

An illustrative example is to apply Eq.~(\ref{eq:bistring}) to the
isosceles triangle configurations in Fourier space such that
\begin{equation}
\left|\bk_1\right| = \left|\bk_2\right|=k,\quad \left|\bk_3\right| = 2
k \sin \dfrac{\theta}{2},
\end{equation}
where $\theta$ denotes the angle between the wavevectors $\bk_1$ and
$\bk_2$. The isosceles bispectrum reads
\begin{equation}
\label{eq:biso}
\BT_{\uiso}(k,\theta) = -\ep^3 \pi
c_0\frac{\vb^2}{\tb^4}\frac{L\corr}{\Area} \dfrac{1}{\corr^2 k^6}\dfrac{1 + 4 \cos
 \theta \sin^2(\theta/2)}{\sin^3\theta}\,.
\end{equation}
\begin{figure}
\begin{center}
\includegraphics[width=9.cm]{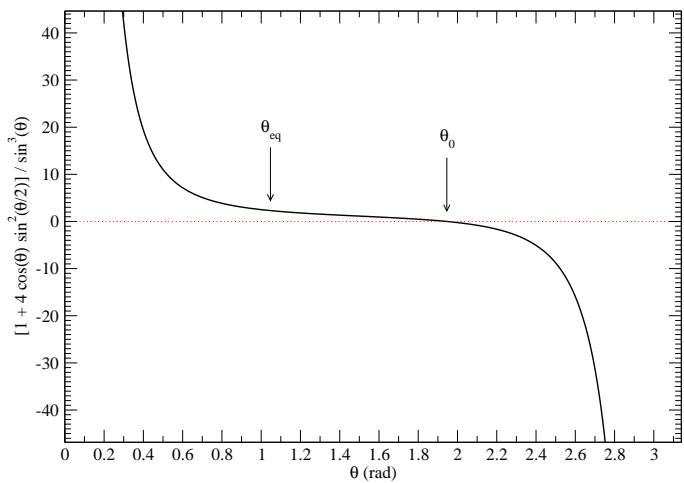}
\caption{Angular dependency of the isosceles bispectrum as a function
  of the angle $\theta$ in between the wavevectors $\bk_1$ and
  $\bk_2$. The particular values $\theta_\ueq = \pi/3$ corresponds to
  the equilateral configuration and $\theta_0$ makes the bispectrum
  vanishing. Notice the amplification for flat triangle configurations
  at $\theta\to0$ (squeezed) and $\theta\to\pi$ (collapsed).}
\label{fig:biso}
\end{center}
\end{figure}

Notice that for $\theta=\pi/3$, we obtain the peculiar case of an
equilateral triangle. In Fig.~\ref{fig:biso}, we have plotted the
angle dependency of the isosceles bispectrum. These configurations are
amplified as $1/\theta^3$ in the two flat triangle limits for which
either $\theta \to 0$ or $\theta \to \pi$. Both of these
configurations are therefore better suited than the equilateral one to
characterize the strings. As suggested by the real space searches, the
strings can produce a strong bispectrum signal only if the detector
resolution is sufficiently good. Assuming a beam resolution of $5'$
means that the $7.2^\circ$ field of view would contain at maximum
roughly $80^2$ Fourier modes. Consequently, the smallest values of
$\angsqz$ achievable would be around $\angsqz > 0.03$ radians, with only
a few modes saturating this bound.

\subsubsection{Numerical results.}

The previous analytical results can be compared to the CMB temperature
bispectrum derived from the simulated maps of
Sec.~\ref{sec:simumaps}. Numerically, one can use the scale
convolution method introduced in Ref.~\cite{Spergel:1999xn,
  Aghanim:2003fs} and applied to the string bispectrum in
Ref.~\cite{Hindmarsh:2009qk}. This method relies on the choice of
unity window functions in Fourier space $\window{u}{\kperp}$ peaked
around a particular wavenumber $u$.  Defining
\begin{equation}
  \Theta_{u}(\vect{x}) \equiv \int\dfrac{\ud \vect{\kperp}}{(2\pi)^2} \FFTheta{\vect{\kperp}}
  \window{u}{\kperp} e^{-i \vect{\kperp} \cdot \vect{x}},
\end{equation}
one can construct an estimator of the three point function in Fourier
space by remarking that
\begin{eqnarray}
  \int \Theta_{k_1}(\vect{x}) \Theta_{k_2}(\vect{x})
  \Theta_{k_3}(\vect{x}) \ud \vect{x} \nonumber \\  = \int \dfrac{\ud \vect{p}\ud
    \vect{q} \ud \vect{k}}{(2\pi)^6}
  \FFTheta{\vect{p}}\FFTheta{\vect{q}} \FFTheta{\vect{k}} 
  \window{k_1}{p}\window{k_2}{q}\window{k_3}{k} (2\pi)^2
  \delta(\vect{p} + \vect{q} + \vect{k}).
\end{eqnarray}
For thin enough window functions, $\FFTheta{\vect{k}}$ remains
constant over the window function width and we construct our reduced
bispectrum estimator as
\begin{equation}
\label{eq:rbisDef}
b_{k_1k_2k_3} = \frac{1}{S^{(w)}_{k_1 k_2 k_3}}
\lvev{\int \Theta_{k_1}(\vect{x})
 \Theta_{k_2}(\vect{x}) \Theta_{k_3}(\vect{x}) \ud \vect{x}}.
\end{equation}
The function $S^{(w)}$ is the flat sky equivalent of the inverse
Wigner-3j symbols and reads
\begin{equation}
\label{eq:invwig}
S^{(w)}_{k_1 k_2 k_3} = \int \dfrac{\ud \vect{p} \ud
 \vect{q}}{(2\pi)^4} \window{k_1}{p} \window{k_2}{q}
\window{k_3}{\left|\vect{p}+\vect{q}\right|}.
\end{equation}
For the window functions such that $\window{u}{k}=1$ for $u-w/2 < k <
u+w/2$, one can approximates
\begin{equation}
\window{u}{k} \simeq w \delta(k-u),
\end{equation}
for small enough values of $w$ compared to the wavenumber $k$. In this
case, Eq.~(\ref{eq:invwig}) can be worked out into
\begin{equation}
\label{eq:appwig}
S^{(w)} \simeq \left( \dfrac{w}{2 \pi}\right)^3 \dfrac{4 k_1 k_2
 k_3}{\sqrt{\left[ \left(k_1 + k_2\right)^2 - k_3^2\right]
   \left[k_3^2 - \left(k_1-k_2\right)^2\right]}}\,.
\end{equation}

\begin{figure}
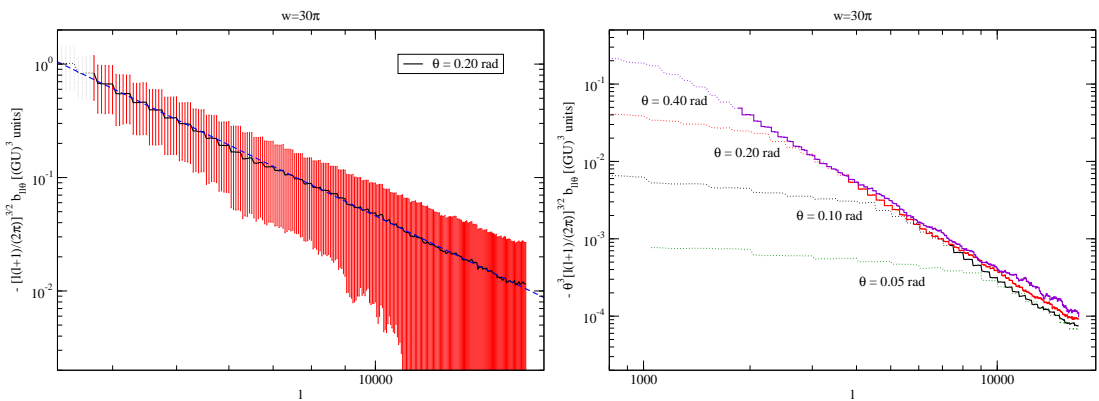

  \begin{center}
    \includegraphics[width=7.2cm]{bllt_hw15pi}
    \includegraphics[width=7.2cm]{blltxtheta3_hw15pi}
    \caption{Left panel: mean value and standard deviation of the
      squeezed isosceles bispectrum
      $\left[\ell(\ell+1)/(2\pi)\right]^{3/2} b_{\ell\ell\angsqz}$ for
      $\angsqz=0.2$ radians. The dashed line is the best power law
      fit. The right panel shows its rescaling by $\angsqz^3
      [\ell(\ell+1)/(2\pi)]^{3/2} b_{\ell\ell\angsqz}$ showing the
      $1/\angsqz^3$ dependency. The spurious plateau (dotted) for the
      lower multipoles comes from a numerical cutoff associated with
      the window functions and occurs at $\ell_{\min}(\angsqz) \simeq
      30\pi/(\angfov \angsqz)$, the field of view being
      $\angfov=7.2^\circ$.}
    \label{fig:bllthw15}
  \end{center}
\end{figure}

In the left panel of Fig.~\ref{fig:bllthw15}, we have plotted the mean
string bispectrum and its standard deviation obtained by this method
over the $300$ string CMB maps. For illustration purpose, this plot is
for the isosceles configuration having $\theta=0.2$ radians. The right
frame of Fig.~\ref{fig:bllthw15} shows the same mean bispectrum but
multiplied by $\theta^3$, for various small values of $\theta$.
As expected from the analytical results, we recover the $1/\theta^3$
behaviour. The wavenumber dependency also matches with the analytical
calculations, up to similar slight power differences as we found for
the power spectrum. A power law fit against the mean numerical
estimator gives
\begin{equation}
\label{eq:blltfit}
\left[\ell(\ell+1)\right]^{3/2} b_{\ell\ell\angsqz} \propto \ell^{-q} \quad \mathrm{with} \quad q \simeq 2.8\,,
\end{equation}
while the overall amplitude can be evaluated around the minimum
variance multipole. At $\ell=5000$, one gets
\begin{equation}
\label{eq:blltamp}
\left. \left[\ell(\ell+1)/(2\pi)\right]^{3/2} b_{\ell\ell\angsqz}\right|_{\ell=5000} \simeq \left(-2.7
  \pm 1.4\right) \times 10^{-3} \left(\dfrac{GU}{\angsqz}\right)^3,
\end{equation}
which also matches with Eq.~(\ref{eq:biso}) under some crude
estimation of the string parameters~\cite{Hindmarsh:2009qk}. Finally,
as suggested by Fig.~\ref{fig:biso}, the string bispectrum is mostly
negative. Integrated over all possible configurations, one recovers
the mean negative sample skewness of Eq.~(\ref{eq:skew}), thereby
explaining its origin as a direct consequence of the breaking of the
time reversal symmetry in FLRW spacetimes.

\subsection{Trispectrum}
\label{sec:trispec}

The trispectrum of the string induced CMB temperature anisotropies can
be derived in a similar way. Starting from the definition of the
four-point functions
\begin{equation}
  \lvev{\FFTheta{\bk_1} \FFTheta{\bk_2}
    \FFTheta{\bk_3} \FFTheta{\bk_4}}  =
  T(\bk_1,\bk_2,\bk_3,\bk_4)(2\pi)^2  \delta(\bk_1+\bk_2+\bk_3+\bk_4),
\end{equation}
we define the trispectrum as\footnote{Notice that our denomination
  ``trispectrum'' here contains the unconnected part. This one is
  however non-vanishing only for parallelogram configurations of the
  wavevectors.}
\begin{eqnarray}
\label{eq:tristart}
&&T(\bk_1,\bk_2,\bk_3,\bk_4) \nonumber \\
&&= \frac{\ep^4}{\Area}
\dfrac{k_{1_A}k_{2_B} k_{3_C} k_{4_D}}
{k_1^2k_2^2k_3^2k_4^2}  \int \ud\sigma_1\ud\sigma_2\ud\sigma_3 \ud\sigma_4
\lvev{\Xd^A_1\Xd^B_2\Xd^C_3 \Xd^D_4 e^{i\delta^{ab}\bk_a \cdot \bX_b}},
\end{eqnarray}
with $\Xd^A_a = \Xd^A(\sigma_a)$, $(a,b) \in \{1,2,3,4\}$ and
$\bk_1+\bk_2+\bk_3+\bk_4=0$. As shown in Ref.~\cite{Hindmarsh:2009es},
the trispectrum and the higher $n$-point functions exhibit
unfactorisable ``flat directions'' in the $n$-dimensional space of the
integration variables $\{\sigma_a\}$. Physically, it means that the
leading order part of the (connected) trispectrum is sensitive to the
higher orders of the correlators in Eqs.~(\ref{eq:velocity}) to
(\ref{eq:tangent}). For the trispectrum, the correlator $T(\sigma)$
has to be expanded at next-to-leading order, and following the
Polchinski--Rocha model~\cite{Polchinski:2006ee}, we assume a
non-analytical behaviour for $T(\sigma)$ at small scales
\begin{equation}
\label{eq:NLOcorr}
  T(\sigma) \simeq \tb^2-\alphac \left(\dfrac{\sigma}{\corr}\right)^{2\chi}.
\end{equation}
In the light-cone gauge, we leave $\alphac$ and $\chi$ as undetermined
parameters since they cannot be straightforwardly inferred from the
numerics performed in the temporal gauge. Nevertheless, because the
correlation should be smaller as $\sigma$ becomes larger, one should
have $\alphac>0$. In the temporal gauge, $\chi$ is directly related to
the power law exponent of the scaling loop distribution functions
through $\chi=1-p/2$. As we are bound to show, the mode dependence of
the trispectrum will also be uniquely given by this parameter. Once
the tangent vector correlator expressed as in Eq.~(\ref{eq:NLOcorr}),
the integrations in Eq.~(\ref{eq:tristart}) can be performed
explicitly, except for parallelogram configurations which have to be
dealt as a special case. After some tedious calculations, an
interpolating expression for the trispectrum
is~\cite{Hindmarsh:2009es}
\begin{eqnarray}
  \label{eq:finaltri}
T(\bk_1,\bk_2,\bk_3,\bk_4) &\simeq& \ep^4 \frac{\vb^4}{\tb^2}
  \frac{L \corr }{\Area} \left(\alphac
    \corr^2\right)^{-1/(2\chi+2)}f(\chi)
\nonumber \\ &\times&
  \gammain{\frac{1}{2\chi+2}}{\betac
    \why^2 \chimax^{2\chi+2}}g(\bk_1,\bk_2,\bk_3,\bk_4).
\end{eqnarray}
In this equation, $f(\chi)$ is a number depending only on the
parameter $\chi$
\begin{equation}
 f(\chi) = \frac{\pi}{\chi+1}\Gamma\left ( \dfrac{1}{2\chi+2} \right) \left[ 4(2\chi+1)(\chi+1)\right]^{1/(2\chi+2)},
\end{equation}
while $g(\{\bk_a\})$ is the trispectrum geometrical factor defined by
\begin{equation}
\label{eq:geometrical}
 g(\bk_1,\bk_2,\bk_3,\bk_4)  = \dfrac{\ka_{12} \ka_{34} + \ka_{13}
    \ka_{24}+ \ka_{14} \ka_{23}}{k_1^2 k_2^2 k_3^2 k_4^2} 
\left [\why^2\right] ^{-1/(2\chi+2)},
\end{equation}
where
\begin{equation}
\why^2(\bk_1,\bk_2,\bk_3,\bk_4) \equiv -\ka_{12} \left(k_3^2 k_4^2 - \kappa_{34}^2\right)^{\chi+1} +
  \circlearrowleft,
\end{equation}
and $\circlearrowleft$ stands for cyclic permutations over the
indices.  The function $\gammain{a}{x}$ stands for the normalised
incomplete lower gamma function defined by
\begin{equation}
\gammain{a}{x} \equiv \dfrac{\gamma(a,x)}{\Gamma(a)} =
\dfrac{1}{\Gamma(a)} \int_0^x t^{a-1} e^{-t} \ud t,
\end{equation}
and, finally, $\Lambda$ has been defined by
\begin{equation}
\label{eq:lambda}
\Lambda(\bk_1,\bk_2,\bk_3,\bk_4) \equiv 
\dfrac{2 L}{\left(k_1^2 k_2^2 -\kappa_{12}^2\right)^{1/2} +
  \circlearrowleft}\times   \dfrac{k_1 k_2 k_3
  k_4}{\kappa_{12}\kappa_{34}+\kappa_{13}\kappa_{24} +
  \kappa_{14}\kappa_{23}} \,.
\end{equation}

As an application, the trispectrum over parallelogram configurations
is obtained when $\why^2=0$ and the leading term of
Eq.~(\ref{eq:finaltri}) simplifies to
\begin{equation}
\label{eq:tripara}
T_0(\bk_1,\bk_2,\bk_3,\bk_4) \simeq
\frac{\pi \ep^4 \vb^4}{\tb^2}\frac{L^2}{\Area k_1^3k_2^3|\sin\theta|}\,,
\end{equation}
where $\theta$ now refers to the parallelogram angle. Under the
scaling transformation $\bk_a \to \lambda \bk_a$, the parallelogram
trispectrum scales as
\begin{equation}
\label{eq:ucscaling}
  T_0(\lambda \bk_1,\lambda \bk_2,\lambda \bk_3,\lambda \bk_4)=\lambda^{-6}T_0(\bk_1,\bk_2,\bk_3,\bk_4).
\end{equation}
For parallelograms, as already mentioned, the trispectrum also gets a
contribution from the unconnected part of the four-point function,
which is Gaussian and reads
\begin{equation}
  T_0^{\uuc}(\bk_1,\bk_2,\bk_3,\bk_4)=\Area P(k_1) P(k_2) +\circlearrowleft.
\end{equation}
Using Eq.~(\ref{eq:powermark}), ones sees that the unconnected part
also behaves as $\lambda^{-6}$. Therefore the non-Gaussian
contributions for parallelogram configurations remain of the same
order of magnitude as the Gaussian ones, with however, and again, an
exception in the squeezed limit $\theta \to 0$.

The most interesting situations come when $\why^2 \neq 0$. For these
quadrilaterals, the Gaussian contribution vanishes and solely a
non-Gaussian statistics can source the trispectrum. At large
wavenumber (small angles), one has $\why^2\gg 1$ such that the
normalised lower incomplete gamma function in Eq.~(\ref{eq:finaltri})
is close to unity:
\begin{equation}
  \label{eq:triinfty}
  T_\infty(\bk_1,\bk_2,\bk_3,\bk_4)\simeq
  \ep^4 \frac{\vb^4}{\tb^2}\frac{L \corr}{\Area} \left(\alphac
    \corr^2\right)^{-1/(2\chi+2)} f(\chi)
  g(\bk_1,\bk_2,\bk_3,\bk_4).
\end{equation}
Under the scaling transformation $\bk_a \to \lambda \bk_a$, the geometric
factor, and thus Eq.~(\ref{eq:triinfty}), scales as
\begin{equation}
  g(\lambda\bk_1,\lambda\bk_2, \lambda\bk_3, \lambda\bk_4) =
  \lambda^{-\rho}
  g(\bk_1,\bk_2,\bk_3,\bk_4),
\end{equation}
with
\begin{equation}
\rho = 6 + \frac{1}{\chi+1}\,.
\end{equation}
As claimed, for NG strings, $\rho$ is directly given by the power
law of the the loop distribution~\cite{Ringeval:2005kr}. Since this
exponent is different from the one associated with parallelogram
configurations it may actually be used to distinguish a trispectrum
sourced by cosmic strings with the one generated by other non-Gaussian
effects.

\begin{figure}
  \begin{center}
    \includegraphics[height=6cm]{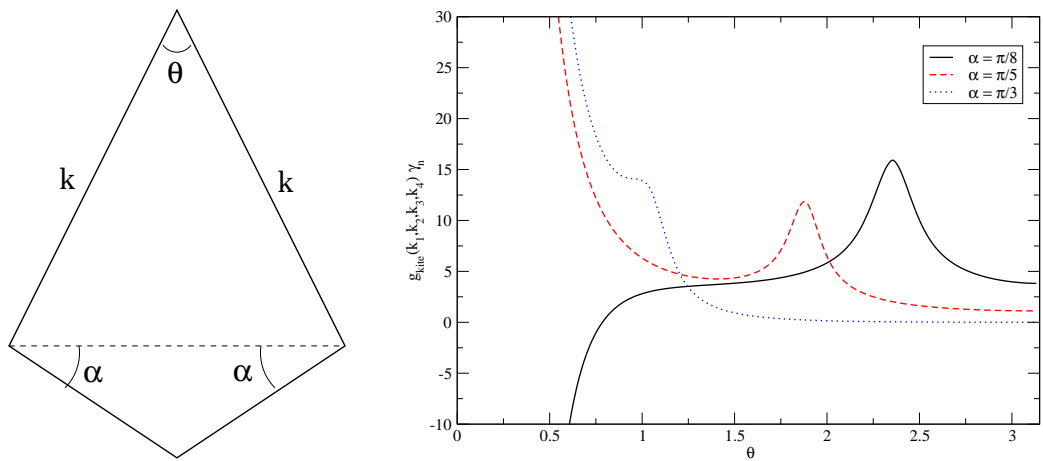}
    \hspace{0.5cm}
    \includegraphics[height=6cm]{triregkite}
    \caption{Trispectrum geometrical factor for the kite
      quadrilaterals (represented on the left panel) as a function of
      the opening angle $\theta$, and plotted for various values of
      $\alpha$. The trispectrum is enhanced in the squeezed limit
      $\theta \to 0$. The bump for $\theta_\up=\pi-2\alpha$
      corresponds to the parallelogram limit for which the unconnected
      part is no longer vanishing.}
    \label{fig:kite}
  \end{center}
\end{figure}

In Fig.~\ref{fig:kite}, we have plotted the geometrical factor
$g(\bk_1,\bk_2,\bk_2,\bk_4)$ for the kite quadrilaterals (represented
in the same figure), as a function of their opening angles $\theta$
and $\alpha$.
As for the bispectrum, the trispectrum is enhanced on the squeezed
configuration obtained when $\theta$ becomes small. In this limit
Eq.~(\ref{eq:geometrical}) can be expanded as
\begin{eqnarray}
  \label{eq:kitesq}
  g & \sim & \dfrac{8 \cos^2(\alpha)}{k^\rho \theta^{\rho-3}} (1-2 \cos
  2\alpha ) \nonumber \\ & \times & \left[ 2(1+\chi) \tan^2(\alpha) - 1 + 4^\chi
    (1-\tan^2 \alpha) \right]^{-1/(2\chi+2)},
\end{eqnarray}
and one recovers the mode dependency in $k^{-\rho}$ while the
amplitude is amplified as $\theta^{\rho-3}$. As discussed in the
previous section, the singular limit $\theta\rightarrow 0$ is never
reached with a finite resolution beam.

\subsection{Comparison with data}

The cosmic string bispectrum and trispectrum associated with the flat
polygonal configurations are the best suited to look for string
signatures. However, it is not easy to compare with existing
constraints as much of the literature focuses on particular models of
\emph{primordial} non-Gaussianity. For instance, in the local type of
primordial non-Gaussianities, the parameter $\fNL$ characterises the
primordial bispectrum and maximal amplitude occurs for squeezed
triangle configurations, as it is the case for the cosmic
strings~\cite{Bartolo:2004if}. However, as a result of the CMB
transfer functions, a given value of $\fNL$ corresponds to oscillating
damped patterns of the CMB \emph{temperature bispectrum}, which are
completely different of the power laws we have found for the string
bispectrum at small scales. The current bounds on $\fNL$ being
precisely obtained from template matching procedures, they cannot be
applied to the strings~\cite{Komatsu:2008hk, Smith:2009jr,
  Regan:2010cn}. For this reason, the parameters used to quantify
primordial non-Gaussianities are not well suited here, precisely
because we expect the string non-Gaussianities to be
non-primordial. An efficient approach would be to use a template
matching procedure with the formulae derived in the previous
sections. Another approach might be to estimate what values the
primordial parameters, such as $\fNL$ and $\tauNL$, would assume if
the non-Gaussianities were actually due to strings. Notice that asking
such a question would be close to find the best amplitude of a sine
function to fit a power law. However, since primordial
non-Gaussianities are and will be tested in CMB data anyway, one could
answer this question by performing a Fisher matrix analysis along the
lines of Refs.~\cite{Sefusatti:2007ih, Nitta:2009jp}.

\section{Conclusion and perspectives}
\label{sec:conc}

The results presented in this article were essentially concerned with
Nambu--Goto type of cosmic strings, which is the simplest realisation
of a one-dimensional spatially extended object. As a result, they
should not be blindly extrapolated to other types of string, although,
as argued in Sec.~\ref{sec:kinds}, some of them are expected to be
generic. In particular, due to the scaling of the long strings, cosmic
string loops do not influence significantly the CMB
observables. Changing the intercommuting probability is expected to
rescale some of the presented results~\cite{Sakellariadou:2004wq,
  Avgoustidis:2007aa}, but in a way which remains to be quantified.

In Sec.~\ref{sec:evol}, we have briefly reviewed the current
understanding of the cosmological evolution of a string network by
means of FLRW numerical simulations, which is a non-trivial problem
even for NG strings. Observable predictions crucially depend on this
step. Numerical simulations can be avoided by making some assumptions
on the string distribution but at the expense of introducing
unnecessary extra parameters. When approximate analytical models are
then used to derive observable predictions, one should keep in mind
that the results are as uncertain as the values assumed for the
additional parameters. Provided one is interested in length scales not
affected by gravitational back-reaction effects, all of the
statistical properties of a NG string network in scaling depends only
on one unknown physical parameter: the string energy density per unit
length $U$, not more\footnote{The expansion rate is supposed to be
  known.}.

In this context, Sec.~\ref{sec:smapng} discusses the non-Gaussian
effects induced by a cosmological string network in the CMB
temperature anisotropies. We have shown that string induced CMB
fluctuations have a negative skewness and a non-vanishing kurtosis. On
a CMB temperature anisotropies map, these non-Gaussianities imprint
characteristic signatures in a multifractal analysis as well as in the
gradient magnitude, both being more significant at small angles. This
property is recovered in Fourier space: the CMB angular power spectrum
decays at most as $1/\ell$, for the large multipoles $\ell$, and
strings become the dominant sources of primary fluctuations. The
skewness appears to be the direct consequence of the breaking of the
time reversal symmetry in an expanding universe, and implies the
existence of a non-vanishing bispectrum. Using analytical
approximations, tested and confirmed by numerical simulations, we then
derived the expected bispectrum and trispectrum of string induced CMB
temperature anisotropies for the large multipoles. Although the
bispectrum decays not faster than $\ell^{-6}$, the trispectrum
multipole dependency is in $\ell^{-\rho}$, where $\rho=6+1/(\chi+1)$
and $\chi$ is a small number related to the tangent vector correlator
and the NG loop distribution. Due to the line-like CMB patterns
induced by the strings, both the bispectrum and trispectrum are
enhanced on all elongated triangle and quadrilateral configurations of
the wavevectors. These ones may constitute the best configurations to
look for a non-Gaussian string signal while being experimentally
limited by finite beam resolution. Let us note that our expressions
have been derived in the flat sky approximation. String
non-Gaussianities at small multipoles is still an open problem which
could be dealt with full sky string maps~\cite{Landriau:2004,
  Regan:2009hv, Landriau:2010cb}. However, if, as the current
constraints suggest, cosmic strings marginally contribute to the large
scale CMB anisotropies, then they should show up at large multipoles
in all of the above mentioned observables. This is precisely where the
experimental efforts are directed. At very small angular scales, the
difficulties will certainly be to separate the string signals from the
astrophysical sources. Interestingly, the very soon accessible
intermediate angles probed by the Planck satellite, and the other
ground based telescopes, may not suffer from this problem and could be
an open window on cosmic strings.

\ack It is a pleasure to thank Patrick Peter, Mairi Sakellariadou,
Daniele Steer and Teruaki Suyama for a careful reading of the
manuscript and their enlightening comments. This work is supported by
the Belgian Federal Office for Scientific, Technical and Cultural
Affairs, under the Inter-University Attraction Pole grant P6/11.

\section*{References}

\bibliography{bibstrings}

\end{document}